\documentclass[pra,twocolumn,longbibliography]{revtex4-1}
\usepackage{graphicx}
\usepackage{amssymb}
\usepackage{bm}
\usepackage{amsmath,amsfonts,latexsym,braket}

\usepackage{hyperref}              
\usepackage{comment}
\usepackage{todonotes}

\usepackage{mathtools}

\newcommand*{\cK}{{\mathcal K}}

\newcommand*{\cH}{{\mathcal H}}

\newcommand*{\cI}{{\mathcal I}}

\newcommand*{\cU}{{\mathcal U}}

\def\beq{\begin{equation}}
\def\eeq{\end{equation}}
\def\bea{\begin{eqnarray}}
\def\eea{\end{eqnarray}}

\DeclareMathOperator{\tr}{tr}
\DeclareMathOperator*{\E}{\mathbb{E}}

\tikzstyle{tensor}=[rectangle,draw=blue!50,fill=blue!20,thick]

\newcommand{\lineH}[3]{\draw (#1,#3) -- (#2,#3);}
\newcommand{\lineV}[3]{\draw (#3,#1) -- (#3,#2);}
\newcommand{\mpsT}[3]{\draw[rounded corners] (0.5+#1,0.5+#2) rectangle (-0.5+#1,-0.5+#2); \draw (#1,#2) node {$#3$};}

\begin{document}

\title{Unitary circuits of finite depth and infinite width from quantum channels}

\author{Sarang Gopalakrishnan}
\affiliation{Department of Physics and Astronomy, CUNY College of Staten Island, Staten Island NY 10314 USA}
\affiliation{Physics Program and Initiative for Theoretical Sciences, The Graduate Center, CUNY, New York NY 10016 USA}

\author{Austen Lamacraft}
\affiliation{TCM Group, Cavendish Laboratory, University of Cambridge,
J. J. Thomson Ave., Cambridge CB3 0HE, UK}

\date{\today}

\begin{abstract}
We introduce an approach to compute reduced density matrices for local quantum unitary circuits of finite depth and infinite width. Suppose the time-evolved state under the circuit is a matrix-product state with bond dimension $D$; then the reduced density matrix of a half-infinite system has the same spectrum as an appropriate $D \times D$ matrix acting on an ancilla space.
We show that reduced density matrices at different spatial cuts are related by quantum channels acting on the ancilla space. This quantum channel approach allows for efficient  numerical evaluation of the entanglement spectrum and R\'enyi entropies and their spatial fluctuations at finite times in an infinite system.
We benchmark our numerical method on random unitary circuits, where many analytic results are available, and also show how our approach analytically recovers the behaviour of the kicked Ising model at the self-dual point.
We study various properties of the spectra of the reduced density matrices and their spatial fluctuations in both the random and translation-invariant cases.
\end{abstract}

\maketitle

\section{Introduction}


The dynamics of isolated quantum systems under generic unitary dynamics is one of the basic problems in many-body physics~\cite{polkovnikov_review}; despite considerable recent work, many aspects of this problem are not fully understood. An isolated system, evolving under chaotic dynamics from an initial product state, becomes increasingly entangled over time. At sufficiently late times, any finite-size subsystem of an infinite system is well described by a thermal reduced density matrix, provided the system obeys the eigenstate thermalization hypothesis (ETH)~\cite{deutsch_eth, srednicki_eth, rigol2008thermalization}; this approach to a thermal local density matrix is called ``thermalization.'' There is considerable numerical evidence that generic many-body systems obey ETH~\cite{deutsch_review}.
%

We are concerned with the dynamics of the reduced density matrix \emph{before} the system has fully thermalized. To quantify the thermalization timescale more precisely, recall that the R\'enyi entropies of a subsystem $A$ are defined in terms of the reduced density matrix $\rho_A$ of $A$ as
\begin{equation}\label{eq:ren_def}
  S^{(n)}_A(t) = \frac{1}{1-n}\log\tr\left[\rho_A(t)^n\right],
\end{equation}
where $n$ is called the R\'enyi index. The R\'enyi entropies fully characterize the spectrum of $\rho_A$.

The general consensus \cite{calabrese2005evolution,Kim:2013aa, nrvh, Keyserlingk2017, Nahum2017} is that -- unless a system experiences many-body localization \cite{Znidaric:2008aa} -- the entropies initially obey $S^{(n)}_A(t)\sim v_n t$, increasing linearly with time with a growth rate $v_n$ that depends on the R\'enyi index $n$. (However, recent results suggest that for $n > 1$ the growth is sub-linear for generic initial states in the presence of conservation laws~\cite{rpv2019, yichen2019}.) The implication for the spectrum of the reduced density matrix is as follows. Parameterizing the eigenvalues of $\rho_A$ in terms of an ``entanglement energy'' as $\lambda_i = e^{-\epsilon_i}$~\cite{ent_spectrum, pollmann2010}, and introducing the ``density of states'' $\varrho(\epsilon)$, we have
\begin{equation}
  S^{(n)}_A = \frac{1}{1-n}\log\left[\int d\epsilon \, \varrho(\epsilon) e^{-n\epsilon} \right].
\end{equation}
The behaviour $S^{(n)}_A(t)\sim v_n t$ is then consistent with a density $\varrho(\epsilon)$ having the large deviation form
\begin{equation}\label{eq:large_dev}
  \varrho(\epsilon) \sim \exp[t \pi(\epsilon/t)],
\end{equation}
for some function $\pi(\eta)$. In the saddle point approximation we find
\begin{equation}\label{eq:rates}
  v_n = \frac{S^{(n)}_A}{t} = \frac{\pi(\eta_n)-\eta_n}{1-n}
\end{equation}
where $\eta_n$ is determined by $n=\pi'(\eta_n)$. The growth rates $v_n$ are seen to be related to the function $p(\eta)$ describing the density of states of the entanglement spectrum by Legendre transformation. Note that the numerator in Eq.~\eqref{eq:rates} vanishes at $n=1$ due to the normalization of the density matrix, yielding a finite growth rate $v_1$ for the von Neumann entropy $S^{(1)}_A$. R\'enyi entropies with smaller $n$ grow faster; the reduced density matrix for a subsystem becomes thermal when the slowest R\'enyi entropy, the so-called ``min-entropy'' $S_\infty \equiv \max_i(\epsilon_i)$ (i.e., the largest entanglement eigenvalue) has saturated.



As well as the rates $v_n$, the time evolution is characterized by the butterfly velocity $v_B$ at which local perturbations spread~\cite{ho2017, mezei2017entanglement, nrvh, Keyserlingk2017, Nahum2017, tianci}.
Although the various velocities are generically separate, they coincide in exactly solvable models (such as random circuits in the limit of large local Hilbert space dimension or Clifford gates~\cite{nrvh, Keyserlingk2017, Nahum2017} and the self-dual kicked Ising model~\cite{Bertini:2018aa, Bertini:2018fbz}), so the entanglement spectrum evolves in a trivial way.
Away from these non-generic limits, little is known analytically about the entanglement spectrum. A few R\'enyi entropies can be explicitly computed by mapping the circuit dynamics to random classical partition functions~\cite{Keyserlingk2017, Nahum2017, tianci} but these mappings do not yield the full entanglement spectrum.
The picture that emerges from numerical studies is, however, that the entanglement spectrum has nontrivial structure in generic systems, such as a bandwidth that widens linearly in time; this feature is absent in the exactly solvable limits~\cite{ccgp}. However, this structure is not fully understood at present. 

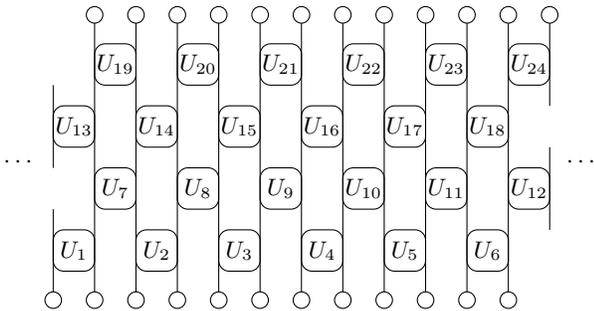
\begin{figure}\label{circfig}
$\cdots$
\begin{tikzpicture}[baseline={([yshift=-1ex]current bounding box.center)},every node/.style={scale=1},scale=.55]
\mpsT{-5}{-3}{U_1}
\mpsT{-3}{-3}{U_2}
\mpsT{-1}{-3}{U_3}
\mpsT{1}{-3}{U_4}
\mpsT{3}{-3}{U_5}
\mpsT{5}{-3}{U_6}

\mpsT{-4}{-1.5}{U_7}
\mpsT{-2}{-1.5}{U_8}
\mpsT{0}{-1.5}{U_9}
\mpsT{2}{-1.5}{U_{10}}
\mpsT{4}{-1.5}{U_{11}}
\mpsT{6}{-1.5}{U_{12}}

\mpsT{-5}{0}{U_{13}}
\mpsT{-3}{0}{U_{14}}
\mpsT{-1}{0}{U_{15}}
\mpsT{1}{0}{U_{16}}
\mpsT{3}{0}{U_{17}}
\mpsT{5}{0}{U_{18}}

\mpsT{-4}{1.5}{U_{19}}
\mpsT{-2}{1.5}{U_{20}}
\mpsT{0}{1.5}{U_{21}}
\mpsT{2}{1.5}{U_{22}}
\mpsT{4}{1.5}{U_{23}}
\mpsT{6}{1.5}{U_{24}}

\lineV{-4}{2.5}{-4.5}
\lineV{-4}{2.5}{-3.5}
\lineV{-4}{2.5}{-2.5}
\lineV{-4}{2.5}{-1.5}
\lineV{-4}{2.5}{-0.5}
\lineV{-4}{2.5}{0.5}
\lineV{-4}{2.5}{1.5}
\lineV{-4}{2.5}{2.5}
\lineV{-4}{2.5}{3.5}
\lineV{-4}{2.5}{4.5}
\lineV{-4}{2.5}{5.5}

\lineV{-4}{-3}{-5.5}
\lineV{-3}{-2}{-5.5}
\lineV{-1}{0}{-5.5}
\lineV{0}{1}{-5.5}

\lineV{-2.5}{-0.5}{6.5}
\lineV{0.5}{2.5}{6.5}

\draw (0.5,-4.2) circle (.2);
\draw (1.5,-4.2) circle (.2);

\draw (2.5,-4.2) circle (.2);
\draw (3.5,-4.2) circle (.2);

\draw (4.5,-4.2) circle (.2);
\draw (5.5,-4.2) circle (.2);

\draw (-1.5,-4.2) circle (.2);
\draw (-0.5,-4.2) circle (.2);

\draw (-2.5,-4.2) circle (.2);
\draw (-3.5,-4.2) circle (.2);

\draw (-4.5,-4.2) circle (.2);
\draw (-5.5,-4.2) circle (.2);

\draw (6.5,2.7) circle (.2);

\draw (0.5,2.7) circle (.2);
\draw (1.5,2.7) circle (.2);

\draw (2.5,2.7) circle (.2);
\draw (3.5,2.7) circle (.2);

\draw (4.5,2.7) circle (.2);
\draw (5.5,2.7) circle (.2);

\draw (-1.5,2.7) circle (.2);
\draw (-0.5,2.7) circle (.2);

\draw (-2.5,2.7) circle (.2);
\draw (-3.5,2.7) circle (.2);

\draw (-4.5,2.7) circle (.2);
\end{tikzpicture}
$\cdots$
\caption{A depth $d=4$ unitary circuit of the type considered in this work.}
\end{figure}

In the present work we develop and apply a numerical transfer-matrix approach to compute the structure of the entanglement spectrum for spatially infinite systems at early times. This approach allows us to access some aspects of entanglement for larger subsystems than were studied in previous work: for instance, we are able to compute the spatial fluctuations of entanglement in systems of $10,000$ sites. Our approach works formally with infinite systems; we assume that the system of interest has been initialized in a product state and then subjected to a finite-depth quantum circuit (i.e., evolution for a finite time) consisting of on-site or nearest-neighbor quantum gates (see Fig.~\ref{circfig}). We compute the spectrum of the reduced density matrix of a bipartition into two semi-infinite regions at an arbitrary point. After applying a quantum circuit of finite depth $t$, the reduced density matrix has rank $q^{t-1}$. We will see that the spectrum of the reduced density matrix can be interpreted as that of a $q^{t-1} \times q^{t-1}$ matrix $R$ acting on an ancilla space. In addition, reduced density matrices across adjacent cuts are related by quantum channels that are straightforward to construct given the circuit. These quantum channels act as transfer matrices for the entanglement spectrum.

The quantum-channel perspective is helpful for a number of reasons. First, using this method one can compute the largest few eigenvalues of the entanglement spectrum for circuits that are too deep to permit direct simulation. Second, as the transfer matrix acts on formally infinite systems, spatial fluctuations of entanglement can be directly studied. Third, standard methods from quantum optics such as the stochastic unraveling of quantum channels~\cite{carmichael_book} can be applied to simulate the dynamics of entanglement on larger scales than direct simulation permits. Finally, for translation-invariant initial-states evolving under translation-invariant circuits, one can iterate the quantum channel until it converges, and thus extract the entanglement spectrum, free of finite-size effects, without incurring the computational overhead of time-evolving a large system.

These are the issues we explore in the present work. Our main results are as follows.
First, we provide algorithms based on quantum channels for computing the entanglement spectrum and its low-energy tail, as well as for computing some R\'enyi entropies, by propagating the quantum channel in an ancilla space.
For the exactly solvable model of Ref.~\cite{Bertini:2018fbz} we analytically demonstrate that the entanglement spectrum is trivial, using the properties of the associated quantum channel.
We compute the behavior of the ``low-energy'' (large Schmidt rank) tail of the entanglement spectrum for random unitary circuits, random circuits with a conservation law, and translation-invariant integrable circuits, acting on various initial states.
%
%
We compute the distributions of the purity and of the min-entropy;
for random circuits we find that both the second R\'enyi entropy and the min-entropy follow Gaussian distributions at the accessible circuit depths (the purity, therefore, follows a log-normal distribution). For circuits with conservation laws or translation-invariant circuits, the nature of this low-entanglement-energy tail is sensitive to the fluctuations in the initial state.
%
We compute the spatial correlations of entanglement, and find that their correlation length grows sub-linearly in time. However, the correlation lengths are short at the accessible times and we are not able to identify a definite exponent.

This paper is organized as follows. In Sec.~\ref{background} we briefly review concepts such as unitary circuits, matrix-product states, and quantum channels, as they apply to the algorithms introduced here. In Sec.~\ref{channeldesc} we describe how to construct quantum channels for unitary circuits, and estimate the complexity of various exact and approximate methods for extracting the entanglement spectrum (or some of its moments). In Sec.~\ref{kickedIsing} we explicitly compute the transfer matrix for the self-dual kicked Ising model, and confirm that all the R\'enyi entropies coincide in this model, as they are known to~\cite{Bertini:2018aa,Bertini:2018fbz}. In Sec.~\ref{results} we present results for the entanglement spectrum, the distributions of purity and min-entropy, and the growth of spatial correlations in the entanglement, in random unitary circuits and some variants of these. Finally Sec.~\ref{conclusions} summarizes our results and discusses future directions.


%

%
%


\section{Background}\label{background}

\subsection{Quantum circuits}

In this work we consider systems that evolve under the application of discrete local unitary gates tiled in the pattern shown in Fig.~\ref{circfig}. Our approach applies both to Floquet systems in which the gates are applied periodically in time, and to random circuits where each gate is drawn independently and randomly. As with the TEBD algorithm~\cite{vidal2007}, it can be extended to continuous time evolution under strictly local Hamiltonians by discretizing the time evolution, e.g., through a Suzuki-Trotter decomposition.

\subsection{Matrix Product States}

In this section we introduce some background material on matrix product states (MPS's), which will form an essential part of the following development. More detailed expositions may be found in Refs.\cite{Perez-Garcia:2006aa,Schollwock:2011aa,Orus:2014aa}.

We consider a quantum system described by $N$ identical subsystems with finite Hilbert space dimension $q$. The (pure) quantum states of the system are therefore defined by vectors in the Hilbert space
\begin{equation}
  \cH_N \equiv \overbrace{\mathbb{C}^q\otimes \cdots \otimes \mathbb{C}^q}^{N\text{ times}}.
\end{equation}
A basis of orthonormal product states has the form
\begin{equation}
  \ket{s_{1:N}} = \ket{s_1}_q \otimes \ket{s_2}_q\cdots \otimes \ket{s_N}_q,
\end{equation}
where $\ket{s}_q$ $s=1,\ldots q$ is an orthonormal basis for $\mathbb{C}^q$, and we have introduced the sequence notation $s_{1:N} \equiv s_1,s_2,\ldots s_N$. By taking components of a vector $\ket{\Psi}\in\cH$
\begin{equation}
  \Psi_{s_{1:N}} = \braket{s_{1:N}|\Psi},
\end{equation}
$\ket{\Psi}$ can be regarded as a rank-$N$ tensor with components $\Psi_{s_{1:N}}$.

An MPS is a tensor of the form
\begin{equation}\label{eq:MPSdef}
  \Psi_{s_{1:N}} = A^{(1)}_{s_1}A^{(2)}_{s_2}\cdots A^{(N)}_{s_N},
\end{equation}
where $A^{(1)}_s, \ldots A^{(N)}_s \in\mathbb{C}^{D_j\times D_{j+1}}$ are matrices, and the numbers $D_j$ $j=1,\ldots N+1$ are known as the \emph{bond dimensions}, with $D_1=D_{N+1}=1$. Thus the product Eq.~\eqref{eq:MPSdef} has the form `row vector, product of matrices, column vector', and yields a complex number for each sequence $s_1:s_N$.

An arbitrary vector $\ket{\Psi}$ may be approximated by an MPS with an error that decreases as $D\equiv\max_j D_j$ increases. We will see that the state arising from applying a unitary circuit to a product state is exactly given by an MPS with $D=q^{d-1}$, where $d$ is the depth of the circuit.

A graphical notation for MPS proves to be extremely convenient, and is discussed extensively in Ref.~\cite{Schollwock:2011aa}. In this notation tensors -- such as the matrices or vectors $A^{(j)}$ -- are represented as boxes, with the number of lines or edges entering a box indicating the number of indices the object bears. An edge joining two vertices indicates the (pairwise) contraction of an index. Thus the MPS in Eq.~\eqref{eq:MPSdef} is denoted
\begin{equation*}
\Psi_{s_{1:N}} =
\begin{tikzpicture}[baseline = (X.base),every node/.style={scale=1},scale=.85]
\draw[rounded corners] (1,2) rectangle (2,1);
\draw (1.5,1.5) node (X) {$A^{(1)}$};
\draw (1.5,.2) node {$s_1$};
\draw (2,1.5) -- (3,1.5);
\draw[rounded corners] (3,2) rectangle (4,1);
\draw (3.5,1.5) node {$A^{(2)}$};
\draw (3.5,.2) node {$s_2$};
\draw (4,1.5) -- (4.5,1.5);
\draw (1.5,1) -- (1.5,.5); \draw (3.5,1) -- (3.5,.5);
\end{tikzpicture} \dots
\begin{tikzpicture}[baseline = (X.base),every node/.style={scale=1},scale=.85]
\draw (0.5,1.5) -- (1,1.5);
\draw[rounded corners] (1,2) rectangle (2,1);
\draw (1.5,1.5) node (X) {$A^{(N)}$};
\draw (1.5,.2) node {$s_N$};
\draw (1.5,1) -- (1.5,.5);
\end{tikzpicture}
\end{equation*}
Here, $A^{(2)}$ has three lines attached because the collection of matrices $A^{(2)}_s\in \mathbb{C}^{D\times D}$ may be regarded as rank-3 tensor $A^{(2)}\in \mathbb{C}^{D\times D\times q}$. The vertical leg represents the indices $s_j$ that live in the `physical space' while the horizontal lines represent indices in the `bond space' (which we shall also refer to as the `ancilla space').

The squared norm of a vector $\ket{\Psi}$ may be calculated by contracting all physical indices between $\Psi_{s_{1:N}}$ and $\bar\Psi_{s_{1:N}}$. This has the graphical representation:
\begin{equation}
  \braket{\Psi|\Psi}=\sum_{s_{1:N}} \bar\Psi_{s_{1:N}}\Psi_{s_{1:N}} =
  \begin{tikzpicture}[anchor=base, baseline,every node/.style={scale=1},scale=.75]
  \draw[rounded corners] (1,1.25) rectangle (2,0.25);
  \draw (1.5,0.5) node (X) {$A^{(1)}$};
  \draw[rounded corners] (1,-0.25) rectangle (2,-1.25);
  \draw (1.5,-1) node (X) {$\bar A^{(1)}$};
  \draw (1.5,0.25) -- (1.5,-0.25);

  \draw (2,0.75) -- (3,0.75);
  \draw (2,-0.75) -- (3,-0.75);

  \draw[rounded corners] (3,1.25) rectangle (4,0.25);
  \draw (3.5,0.5) node (X) {$A^{(2)}$};
  \draw[rounded corners] (3,-0.25) rectangle (4,-1.25);
  \draw (3.5,-1) node (X) {$\bar A^{(2)}$};
  \draw (3.5,0.25) -- (3.5,-0.25);

  \draw (4,0.75) -- (4.5,0.75);
  \draw (4,-0.75) -- (4.5,-0.75);

  \end{tikzpicture} \dots
  \begin{tikzpicture}[anchor=base, baseline,every node/.style={scale=1},scale=.75]
  \draw (0.5,0.75) -- (1,0.75);
  \draw (0.5,-0.75) -- (1,-0.75);

  \draw[rounded corners] (1,1.25) rectangle (2,0.25);
  \draw (1.5,0.5) node (X) {$A^{(N)}$};
  \draw[rounded corners] (1,-0.25) rectangle (2,-1.25);
  \draw (1.5,-1) node (X) {$\bar A^{(N)}$};
  \draw (1.5,0.25) -- (1.5,-0.25);
  \end{tikzpicture}
\end{equation}

More generally, the reduced density matrix for the leftmost $n$ subsystems that arises from a pure state by tracing over the remaining subsystems is denoted
\begin{widetext}
  \begin{equation}\label{eq:rdm}
    \rho_{s_{1:n},s'_{1:n}} \equiv\sum_{s_{n+1:N}} \bar\Psi_{s_{1:n}s_{n+1:N}}\Psi_{s'_{1:n}s_{n+1:N}}=
    \begin{tikzpicture}[anchor=base, baseline,every node/.style={scale=0.9},scale=.75]
    \draw[rounded corners] (1,1.75) rectangle (2,0.75);
    \draw (1.5,1) node (X) {$1$};
    \draw[rounded corners] (1,-0.75) rectangle (2,-1.75);
    \draw (1.5,-1.5) node (X) {$\bar 1$};

    \draw (1.5,0.75) -- (1.5,0.4);
    \draw (1.5,0.2) node {$s_1$};

    \draw (1.5,-0.75) -- (1.5,-0.4);
    \draw (1.5,-0.3) node {$s'_1$};

    \draw (2,1.25) -- (2.5,1.25);
    \draw (2,-1.25) -- (2.5,-1.25);

    \end{tikzpicture} \dots
    \begin{tikzpicture}[anchor=base, baseline,every node/.style={scale=0.9},scale=.75]

    \draw (0.5,1.25) -- (1,1.25);
    \draw (0.5,-1.25) -- (1,-1.25);

    \draw[rounded corners] (1,1.75) rectangle (2,0.75);
    \draw (1.5,1) node (X) {$n$};
    \draw[rounded corners] (1,-0.75) rectangle (2,-1.75);
    \draw (1.5,-1.5) node (X) {$\bar n$};

    \draw (1.5,0.75) -- (1.5,0.4);
    \draw (1.5,0.2) node {$s_n$};

    \draw (1.5,-0.75) -- (1.5,-0.4);
    \draw (1.5,-0.3) node {$s'_n$};

    \draw (2,1.25) -- (3,1.25);
    \draw (2,-1.25) -- (3,-1.25);

    \draw[rounded corners] (3,1.75) rectangle (4,0.75);
    \draw (3.5,1) node (X) {$n+1$};
    \draw[rounded corners] (3,-0.75) rectangle (4,-1.75);
    \draw (3.5,-1.5) node (X) {$\overline {n+1}$};

    \draw (3.5,0.75) -- (3.5,-0.75);

    \draw (4,1.25) -- (4.5,1.25);
    \draw (4,-1.25) -- (4.5,-1.25);

    \end{tikzpicture}\cdots
    \begin{tikzpicture}[anchor=base, baseline,every node/.style={scale=0.9},scale=.75]

      \draw (0.5,1.25) -- (1,1.25);
      \draw (0.5,-1.25) -- (1,-1.25);

    \draw[rounded corners] (1,1.75) rectangle (2,0.75);
    \draw (1.5,1) node (X) {$N$};
    \draw[rounded corners] (1,-0.75) rectangle (2,-1.75);
    \draw (1.5,-1.5) node (X) {$\bar N$};

    \draw (1.5,0.75) -- (1.5,-0.75);
    \end{tikzpicture},
  \end{equation}
  \end{widetext}
where for simplicity we denote $A^{(j)}$ by $j$ and $a'^{(j)}$ by $\bar j$.  The above expressions may also be written
\begin{equation}\label{eq:rdm_R}
  \rho_{s_{1:n},s'_{1:n}} =
  \begin{tikzpicture}[anchor=base, baseline,every node/.style={scale=0.9},scale=.75]
  \draw[rounded corners] (1,1.75) rectangle (2,0.75);
  \draw (1.5,1) node (X) {$1$};
  \draw[rounded corners] (1,-0.75) rectangle (2,-1.75);
  \draw (1.5,-1.5) node (X) {$\bar 1$};

  \draw (1.5,0.75) -- (1.5,0.4);
  \draw (1.5,0.2) node {$s_1$};

  \draw (1.5,-0.75) -- (1.5,-0.4);
  \draw (1.5,-0.3) node {$s'_1$};

  \draw (2,1.25) -- (2.5,1.25);
  \draw (2,-1.25) -- (2.5,-1.25);

  \end{tikzpicture} \dots
  \begin{tikzpicture}[anchor=base, baseline,every node/.style={scale=0.9},scale=.75]

  \draw (0.5,1.25) -- (1,1.25);
  \draw (0.5,-1.25) -- (1,-1.25);

  \draw[rounded corners] (1,1.75) rectangle (2,0.75);
  \draw (1.5,1) node (X) {$n$};
  \draw[rounded corners] (1,-0.75) rectangle (2,-1.75);
  \draw (1.5,-1.5) node (X) {$\bar n$};

  \draw (1.5,0.75) -- (1.5,0.4);
  \draw (1.5,0.2) node {$s_n$};

  \draw (1.5,-0.75) -- (1.5,-0.4);
  \draw (1.5,-0.3) node {$s'_n$};

  \draw (3,0) circle (.5);
  \draw (3,-0.1) node (X) {$R^{(n)}$};
  \draw (3,0.5) edge[out=90,in=0] (2,1.25);
  \draw (3,-0.5) edge[out=270,in=0] (2,-1.25);

  \end{tikzpicture},
\end{equation}
where the hermitian matrices $R^{(j)}\in \mathbb{C}^{D_{j+1}\times D_{j+1}}$ are defined by
\begin{equation}\label{eq:CPTP}
R^{(j-1)} = \sum_s A^{(j)\vphantom{\dagger}}_{s} R^{(j)} A^{(j)\dagger}_{s}, \qquad j=n+1,\ldots N,
\end{equation}
and $R^{(N)}=1$.

\subsubsection{Canonical Forms}

The MPS representation of $\ket{\Psi}$ has a redundancy sometimes referred to as `gauge freedom'. For a set $X_j$ $j=1,\ldots N-1$ of invertible matrices, the transformation
\begin{align}
  A^{(1)}_s &\to A^{(1)}_s X_{1}^{-1},\nonumber\\
  A^{(j)}_s &\to X_{j-1}A^{(j)}_s X_{j}^{-1},\qquad j=2,\ldots N-1\nonumber\\
  A^{(N)}_s &\to X_{N-1}A^{(N)}_s ,
\end{align}
leaves $\Psi_{s_{1:N}}$ unchanged. Further conditions may be imposed to reduce this redundancy \cite{Perez-Garcia:2006aa}. We will be concerned with matrices in \emph{left canonical} form, satisfying the condition
\begin{align}\label{eq:left-c}
  \sum_s A^{(j)\dagger}_{s} A^{(j)\vphantom{\dagger}}_{s} &= \openone_{D_{j+1}}\\
 \begin{tikzpicture}[baseline={([yshift=-1ex]current bounding box.center)},every node/.style={scale=0.75},scale=.55]
 \draw (1,-1.5) edge[out=180,in=180] (1,1.5);
 \draw[rounded corners] (1,2) rectangle (2,1);
 \draw[rounded corners] (1,-1) rectangle (2,-2);
 \draw (1.5,1) -- (1.5,-1);
 \draw (1.5,1.5) node {$A^{(j)}$};
 \draw (1.5,-1.5) node {$\bar{A}^{(j)}$};
 \draw (2,1.5) -- (2.5,1.5); \draw (2,-1.5) -- (2.5,-1.5);
 \end{tikzpicture}
 &=
 \begin{tikzpicture}[baseline={([yshift=-1ex]current bounding box.center)},every node/.style={scale=0.750},scale=.55]
 \draw (1,-1.5) edge[out=180,in=180] (1,1.5);
 \end{tikzpicture}.
\end{align}
Note that $D_2=q$ is necessary for $A^{(1)}$ to be placed in left canonical form, but this places no restriction on the state.

Analogously, matrices in \emph{right} canonical form satisfy
\begin{align}\label{eq:right-c}
  \sum_s A^{(j)\vphantom{\dagger}}_{s}A^{(j)\dagger}_{s} &= \openone_{D_{j}}\\
 \begin{tikzpicture}[baseline={([yshift=-1ex]current bounding box.center)},every node/.style={scale=0.75},scale=.55]
 \draw (2,-1.5) edge[out=0,in=0] (2,1.5);
 \draw[rounded corners] (1,2) rectangle (2,1);
 \draw[rounded corners] (1,-1) rectangle (2,-2);
 \draw (1.5,1) -- (1.5,-1);
 \draw (1.5,1.5) node {$A^{(j)}$};
 \draw (1.5,-1.5) node {$\bar{A}^{(j)}$};
 \draw (0.5,1.5) -- (1,1.5); \draw (0.5,-1.5) -- (1,-1.5);
 \end{tikzpicture}
 &=
 \begin{tikzpicture}[baseline={([yshift=-1ex]current bounding box.center)},every node/.style={scale=0.750},scale=.55]
 \draw (1,-1.5) edge[out=0,in=0] (1,1.5);
 \end{tikzpicture}.
\end{align}
(requiring $D_N=q$).

One benefit of the canonical forms is that they may be contracted ``automatically''. For example, in terms of an MPS in right canonical form the reduced density matrix $\rho_{s_{1:n},s'_{1:n}}$ in Eq.~\eqref{eq:rdm} takes the form
\begin{equation}\label{eq:rdm_canon}
  \rho_{s_{1:n},s'_{1:n}} =
  \begin{tikzpicture}[anchor=base, baseline,every node/.style={scale=0.9},scale=.75]
  \draw[rounded corners] (1,1.75) rectangle (2,0.75);
  \draw (1.5,1) node (X) {$1$};
  \draw[rounded corners] (1,-0.75) rectangle (2,-1.75);
  \draw (1.5,-1.5) node (X) {$\bar 1$};

  \draw (1.5,0.75) -- (1.5,0.4);
  \draw (1.5,0.2) node {$s_1$};

  \draw (1.5,-0.75) -- (1.5,-0.4);
  \draw (1.5,-0.3) node {$s'_1$};

  \draw (2,1.25) -- (2.5,1.25);
  \draw (2,-1.25) -- (2.5,-1.25);

  \end{tikzpicture} \dots
  \begin{tikzpicture}[anchor=base, baseline,every node/.style={scale=0.9},scale=.75]

  \draw (0.5,1.25) -- (1,1.25);
  \draw (0.5,-1.25) -- (1,-1.25);

  \draw[rounded corners] (1,1.75) rectangle (2,0.75);
  \draw (1.5,1) node (X) {$n$};
  \draw[rounded corners] (1,-0.75) rectangle (2,-1.75);
  \draw (1.5,-1.5) node (X) {$\bar n$};

  \draw (1.5,0.75) -- (1.5,0.4);
  \draw (1.5,0.2) node {$s_n$};

  \draw (1.5,-0.75) -- (1.5,-0.4);
  \draw (1.5,-0.3) node {$s'_n$};

  \draw (2,-1.25) edge[out=0,in=0] (2,1.25);

  \end{tikzpicture}.
\end{equation}
A second benefit -- which is more relevant for us -- is that the spectrum of the reduced density matrix $\rho_{s_{1:n},s'_{1:n}}$ coincides with the spectrum of $R^{(n)}$ for an MPS in \emph{left} canonical form. This may be seen by introducing the spectral representation
\begin{equation}
  R^{(n)} = \sum_\alpha \lambda_\alpha r_\alpha r_\alpha^\dagger,
\end{equation}
in terms of the eigenvalues $\lambda_\alpha$ and eigenvectors $r_\alpha$ of $R^{(n)}$. Upon substitution into Eq.~\eqref{eq:rdm_R} this yields a spectral representation for $\rho_{s_{1:n},s'_{1:n}}$
\begin{equation}\label{eq:rdm_spectrum}
  \rho_{s_{1:n},s'_{1:n}} = \sum_\alpha \lambda_\alpha
  \begin{tikzpicture}[anchor=base, baseline,every node/.style={scale=0.9},scale=.75]
  \draw[rounded corners] (1,1.75) rectangle (2,0.75);
  \draw (1.5,1) node (X) {$1$};
  \draw[rounded corners] (1,-0.75) rectangle (2,-1.75);
  \draw (1.5,-1.5) node (X) {$\bar 1$};

  \draw (1.5,0.75) -- (1.5,0.4);
  \draw (1.5,0.2) node {$s_1$};

  \draw (1.5,-0.75) -- (1.5,-0.4);
  \draw (1.5,-0.3) node {$s'_1$};

  \draw (2,1.25) -- (2.5,1.25);
  \draw (2,-1.25) -- (2.5,-1.25);

  \end{tikzpicture} \dots
  \begin{tikzpicture}[anchor=base, baseline,every node/.style={scale=0.9},scale=.75]

  \draw (0.5,1.25) -- (1,1.25);
  \draw (0.5,-1.25) -- (1,-1.25);

  \draw[rounded corners] (1,1.75) rectangle (2,0.75);
  \draw (1.5,1) node (X) {$n$};
  \draw[rounded corners] (1,-0.75) rectangle (2,-1.75);
  \draw (1.5,-1.5) node (X) {$\bar n$};

  \draw (1.5,0.75) -- (1.5,0.4);
  \draw (1.5,0.2) node {$s_n$};

  \draw (1.5,-0.75) -- (1.5,-0.4);
  \draw (1.5,-0.3) node {$s'_n$};

  \draw (2,1.25) -- (2.5,1.25);
  \draw (2,-1.25) -- (2.5,-1.25);

  \draw (3,1.25) circle (.5);
  \draw (3,1.1) node (X) {$r_\alpha$};

  \draw (3,-1.25) circle (.5);
  \draw (3,-1.3) node (X) {$\bar r_\alpha$};

  \end{tikzpicture},
\end{equation}
in terms of vectors in $\cH_n$ that are orthonormal by the left canonical condition \cite{Perez-Garcia:2006aa}.

\begin{figure}[!b]
\begin{center}
\includegraphics[width = 0.45\textwidth]{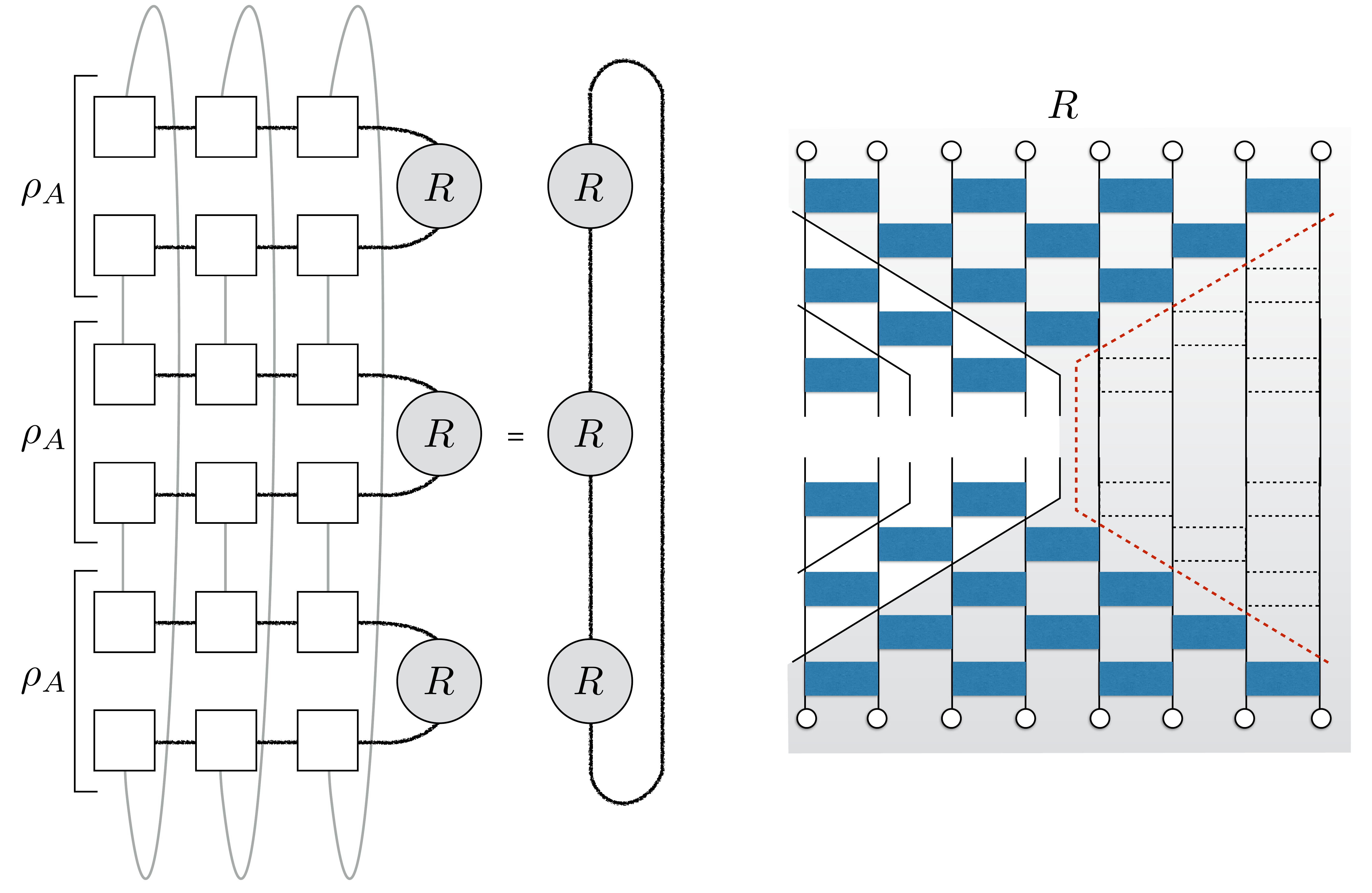}
\caption{Left: the reduced density matrix $\rho_A$ can be expressed as a matrix product operator involving matrices in left canonical form (squares), and a matrix $R$ that comes from the complement, $\bar A$. R\'enyi entropies are proportional to $\tr \rho_A^n$; when the MPS is in left canonical form, $\tr \rho_A^n = \tr R^n$. Right: application of this idea to a unitary circuit. The matrix-product state $|\psi(t)\rangle$ is constructed by cutting diagonally through the circuit (solid lines); the leftover gates in the shaded region form the matrix $R$.}
\label{cforms}
\end{center}
\end{figure}

A more direct way of seeing that the spectra of $R^{(n)}$ and $\rho_A$ coincide, for MPS's in left canonical form, is shown in Fig.~\ref{cforms}. The moments of the spectrum are given by matrix products of the form $\rho_A^k$; one can use the left canonical property~\eqref{eq:left-c} to eliminate the ``boxes'' pairwise and arrive at the result $\tr \rho_A^k = \tr[ (R^{(n)})^k]$. Since all moments coincide, $\rho_A$ and $R^{(n)}$ must have the same spectrum.

\subsection{Quantum Channels}

The above formalism has a natural interpretation in terms of quantum channels, or completely positive trace preserving (CPTP) maps \cite{Perez-Garcia:2006aa}. In this interpretation the matrices $R^{(j)}$ are regarded as a sequence of density matrices that represent mixed states in the bond space. The definition of $R^{(j)}$ given in Eq.~\eqref{eq:CPTP} guarantees that the maps from one $R^{(j)}$ to the next are completely positive (Choi's theorem), while the left canonical condition Eq.\eqref{eq:left-c} ensures that they are trace preserving. Therefore, contracting a physical leg (i.e., moving the entanglement cut in real space) amounts to applying a CPTP map to the ancilla. In this context the $A^{(j)}_s$ are known as Kraus operators.

From now on we assume without loss of generality that all $D_j=D$, which may be achieved by padding the matrices with zeros. Square matrices $A^{(j)}_s\in \mathbb{C}^{D\times D}$ satisfying the left canonical condition may be parameterized in terms of unitary matrices $U_j\in\cU(qD)$ as
\begin{equation}\label{eq:U_ensemble}
A^{(j)}_{s,ab} = \bra{s}_q\bra{a}_D U_j\ket{0}_q\ket{b}_D.
\end{equation}
Physically, this corresponds to the amplitude for the following process: prepare the physical subsystem $j$ in a fixed state $\ket{0}_q$, and then act on the state $\ket{0}_q\ket{b}_D$ of the subsystem $j$ and ancilla with a unitary, arriving in state $\ket{s}_q\ket{a}_D$. While any CPTP map may be presented in this form \cite{lindblad1976generators}, we will see that this is precisely how quantum channels arise in the case of unitary circuits.

\section{Quantum channels from unitary circuits}\label{channeldesc}

This section is organized as follows. We first discuss how to slice up a unitary circuit into an MPS where all matrices are in canonical form. This immediately gives us a quantum channel that propagates the ancilla-space density matrix $R$ in the spatial direction. We then discuss two controlled but approximate ways of implementing the quantum channel for larger subsystems: first, an approach for computing the entanglement spectrum based on approximating the ancilla-space density matrix $R$ as a lower-rank object; and second, an approach for computing the purity, specifically, by a stochastic unraveling of the quantum channel (i.e., by sampling ``quantum trajectories'' in ancilla space~\cite{carmichael_book}).

%
%
%
%
%
%
%
%
%

\subsection{From unitary circuits to canonical-form MPS's}

We now introduce the main idea behind our approach. A planar unitary circuit that starts from a product state may be presented as an MPS by slicing it into strips in an arbitrary way.
Each slice $j$ is associated with $q^2$ matrices $A^{(j)}_{s_1,s_2}$ indexed by two physical indices. The dimension of the ancilla is $D=q^{d-1}$.
%
For a general decomposition of a circuit, the resulting matrices $A^{(j)}_{s_1,s_2}$ will not be in the appropriate canonical form, so the spectra of $R^{(j)}$ and $\rho_A$ will not coincide.
%
%
An MPS in canonical form \emph{is} obtained for slices along the south-west to north-east diagonal
\begin{equation}A^{(j)}_{s_1,s_2;a_{1:3},b_{1:3}}=
\begin{tikzpicture}[baseline={([yshift=-1ex]current bounding box.center)},every node/.style={scale=1},scale=.55]

\mpsT{1}{-3}{U_1}
\mpsT{2}{-1.5}{U_{2}}
\mpsT{3}{0}{U_{3}}
\mpsT{4}{1.5}{U_{4}}

\draw (0.5,-2.75) edge[out=90,in=0] (0,-2.25);
\draw (1.5,-1.25) edge[out=90,in=0] (1,-0.75);
\draw (2.5,0.25) edge[out=90,in=0] (2,0.75);

\draw (2.5,-1.75) edge[out=-90,in=180] (3,-2.25);
\draw (3.5,-0.25) edge[out=-90,in=180] (4,-0.75);
\draw (4.5,1.25) edge[out=-90,in=180] (5,0.75);

\lineV{-4}{-3}{0.5}

\lineV{-4}{-1.5}{1.5}
\lineV{-1.5}{0}{2.5}
\lineV{0}{2.5}{3.5}
\lineV{1.5}{2.5}{4.5}

\draw (0.5,-4.2) circle (.2);
\draw (1.5,-4.2) circle (.2);

\draw (3.5,2.7) node (X) {$s_1$};
\draw (4.5,2.7) node (X) {$s_2$};

\draw (-0.3,-2.25) node (X) {$a_3$};
\draw (0.7,-0.75) node (X) {$a_2$};
\draw (1.7,0.75) node (X) {$a_1$};

\draw (3.3,-2.25) node (X) {$b_3$};
\draw (4.3,-0.75) node (X) {$b_2$};
\draw (5.3,0.75) node (X) {$b_1$};

\end{tikzpicture}.
\end{equation}
A graphical proof is straightforward, since
\begin{equation}\sum_{s_1,s_2} \left(A^{(j)\dagger}_{s_1,s_2} A^{(j)}_{s_1,s_2}\right)_{b'_{1:3},b_{1:3}}=
\begin{tikzpicture}[baseline={([yshift=-1ex]current bounding box.center)},every node/.style={scale=1},scale=.55]

\mpsT{1}{-3}{U_1}
\mpsT{2}{-1.5}{U_{2}}
\mpsT{3}{0}{U_{3}}
\mpsT{4}{1.5}{U_{4}}

\mpsT{1}{7.5}{U^\dagger_1}
\mpsT{2}{6}{U^\dagger_{2}}
\mpsT{3}{4.5}{U^\dagger_{3}}
\mpsT{4}{3}{U^\dagger_{4}}

\draw (2.5,-1.75) edge[out=-90,in=180] (3,-2.25);
\draw (3.5,-0.25) edge[out=-90,in=180] (4,-0.75);
\draw (4.5,1.25) edge[out=-90,in=180] (5,0.75);

\draw (2.5,6.25) edge[out=90,in=180] (3,6.75);
\draw (3.5,4.75) edge[out=90,in=180] (4,5.25);
\draw (4.5,3.25) edge[out=90,in=180] (5,3.75);

\lineV{-4}{8.5}{0.5}
\lineV{-4}{8.5}{1.5}

\lineV{-1.5}{6}{2.5}
\lineV{0}{4.5}{3.5}
\lineV{1.5}{3}{4.5}

\draw (0.5,-4.2) circle (.2);
\draw (1.5,-4.2) circle (.2);

\draw (0.5,8.7) circle (.2);
\draw (1.5,8.7) circle (.2);

\draw (3.3,-2.25) node (X) {$b_3$};
\draw (4.3,-0.75) node (X) {$b_2$};
\draw (5.3,0.75) node (X) {$b_1$};

\draw (3.3,6.75) node (X) {$b'_3$};
\draw (4.3,5.25) node (X) {$b'_2$};
\draw (5.3,3.75) node (X) {$b'_1$};

\end{tikzpicture}.
\end{equation}
The unitarity of the constituent blocks
\begin{equation}
\begin{tikzpicture}[baseline={([yshift=-1ex]current bounding box.center)},every node/.style={scale=1},scale=.55]

\mpsT{1}{0}{U_1}
\mpsT{1}{1.5}{U^\dagger_1}
\lineV{-1}{2.5}{0.5}
\lineV{-1}{2.5}{1.5}

\draw (0.5,-1.5) node (X) {$b_1$};
\draw (1.5,-1.5) node (X) {$b_2$};

\draw (0.5,3) node (X) {$b'_1$};
\draw (1.5,3) node (X) {$b'_2$};

\end{tikzpicture}=
\begin{tikzpicture}[baseline={([yshift=-1ex]current bounding box.center)},every node/.style={scale=1},scale=.55]

\lineV{-1}{2.5}{0.5}
\lineV{-1}{2.5}{1.5}

\draw (0.5,-1.5) node (X) {$b_1$};
\draw (1.5,-1.5) node (X) {$b_2$};

\draw (0.5,3) node (X) {$b'_1$};
\draw (1.5,3) node (X) {$b'_2$};

\end{tikzpicture}
=\delta_{b_1,b'_1}\delta_{b_2,b'_2}
\end{equation}
guarantees that the left canonical condition
\begin{equation}
  \sum_{s_1,s_2} A^{(j)\dagger}_{s_1,s_2} A^{(j)\vphantom{\dagger}}_{s_1,s_2} = \openone_{D}
\end{equation}
is satisfied. Choosing the south-east to north-west diagonal gives an MPS in right canonical form.

\subsection{Applying the quantum channel}

We have shown that a unitary circuit may be represented as an MPS in canonical form. This observation may be used to efficiently compute the ancilla density matrices $R^{(j)}$, whose spectrum coincides with that of the reduced density matrix $\rho_{s_{1:j},s'_{1:j}}$.

Recall that the $R^{(j)}$ are defined by
\begin{equation}
R^{(j-1)} = \sum_{s_1,s_2} A^{(j)\vphantom{\dagger}}_{s_1,s_2} R^{(j)} A^{(j)\dagger}_{s_1,s_2}, \qquad j=n+1,\ldots N.
\end{equation}
After cutting along the SW-NE diagonal, $R^{(j-1)}_{a_{1:d-1},a'_{1:d-1}}$  has the graphical representation
\begin{equation}R^{(j-1)}_{a_{1:d-1},a'_{1:d-1}}=
\begin{tikzpicture}[baseline={([yshift=-1ex]current bounding box.center)},every node/.style={scale=1},scale=.55]

\mpsT{1}{-3}{U_1}
\mpsT{2}{-1.5}{U_{2}}
\mpsT{3}{0}{U_{3}}
\mpsT{4}{1.5}{U_{4}}

\mpsT{1}{7.5}{U^\dagger_1}
\mpsT{2}{6}{U^\dagger_{2}}
\mpsT{3}{4.5}{U^\dagger_{3}}
\mpsT{4}{3}{U^\dagger_{4}}

\draw (2.5,-1.75) edge[out=-90,in=180] (3,-2.25);
\draw (3.5,-0.25) edge[out=-90,in=180] (4,-0.75);
\draw (4.5,1.25) edge[out=-90,in=180] (5,0.75);

\draw (2.5,6.25) edge[out=90,in=180] (3,6.75);
\draw (3.5,4.75) edge[out=90,in=180] (4,5.25);
\draw (4.5,3.25) edge[out=90,in=180] (5,3.75);

\draw (0.5,-2.75) edge[out=90,in=0] (0,-2.25);
\draw (1.5,-1.25) edge[out=90,in=0] (1,-0.75);
\draw (2.5,0.25) edge[out=90,in=0] (2,0.75);

\draw (0.5,7.25) edge[out=-90,in=0] (0,6.75);
\draw (1.5,5.75) edge[out=-90,in=0] (1,5.25);
\draw (2.5,4.25) edge[out=-90,in=0] (2,3.75);

\lineV{-4}{-3}{0.5}
\lineV{-4}{-1.5}{1.5}

\lineV{7.5}{8.5}{0.5}
\lineV{6}{8.5}{1.5}

\lineV{-1.5}{0}{2.5}
\lineV{4.5}{6}{2.5}

\lineV{0}{4.5}{3.5}
\lineV{1.5}{3}{4.5}

\draw (0.5,-4.2) circle (.2);
\draw (1.5,-4.2) circle (.2);

\draw (0.5,8.7) circle (.2);
\draw (1.5,8.7) circle (.2);

%

\draw (-0.3,-2.25) node (X) {$a_3$};
\draw (0.7,-0.75) node (X) {$a_2$};
\draw (1.7,0.75) node (X) {$a_1$};

\draw (1.7,3.75) node (X) {$a'_1$};
\draw (0.7,5.25) node (X) {$a'_2$};
\draw (-0.3,6.75) node (X) {$a'_3$};

\draw (7,2.25) node (X) {$R^{(j)}$};
\draw[rounded corners] (6,3) rectangle (8,1.5);

\draw (5,0.75) edge[out=0,in=-90] (6.5,1.5);
\draw (4,-0.75) edge[out=0,in=-90] (7,1.5);
\draw (3,-2.25) edge[out=0,in=-90] (7.5,1.5);

\draw (5,3.75) edge[out=0,in=90] (6.5,3);
\draw (4,5.25) edge[out=0,in=90] (7,3);
\draw (3,6.75) edge[out=0,in=90] (7.5,3);

\end{tikzpicture}.
\end{equation}
This expression may be simplified somewhat by noting that the topmost unitary ($U_4$ in the above example) may be eliminated to give
\begin{equation}R^{(j-1)}_{a_{1:d-1},a'_{1:d-1}}=
\begin{tikzpicture}[baseline={([yshift=-1ex]current bounding box.center)},every node/.style={scale=1},scale=.55]

\mpsT{1}{-2}{U_1}
\mpsT{2}{-0.5}{U_{2}}
\mpsT{3}{1}{U_{3}}

\mpsT{1}{6.5}{U^\dagger_1}
\mpsT{2}{5}{U^\dagger_{2}}
\mpsT{3}{3.5}{U^\dagger_{3}}

\draw (2.5,-0.75) edge[out=-90,in=180] (3,-1.25);
\draw (3.5,0.75) edge[out=-90,in=180] (4,0.25);
\draw (4.5,2.25) edge[out=-90,in=180] (5,1.75);

\draw (2.5,5.25) edge[out=90,in=180] (3,5.75);
\draw (3.5,3.75) edge[out=90,in=180] (4,4.25);
\draw (4.5,2.25) edge[out=90,in=180] (5,2.75);

\draw (0.5,-1.75) edge[out=90,in=0] (0,-1.25);
\draw (1.5,-0.25) edge[out=90,in=0] (1,0.25);
\draw (2.5,1.25) edge[out=90,in=0] (2,1.75);

\draw (0.5,6.25) edge[out=-90,in=0] (0,5.75);
\draw (1.5,4.75) edge[out=-90,in=0] (1,4.25);
\draw (2.5,3.25) edge[out=-90,in=0] (2,2.75);

\lineV{-3}{-2}{0.5}
\lineV{-3}{-0.5}{1.5}

\lineV{6.5}{7.5}{0.5}
\lineV{5}{7.5}{1.5}

\lineV{-0.5}{1}{2.5}
\lineV{3.5}{5}{2.5}

\lineV{1}{3.5}{3.5}

\draw (0.5,-3.2) circle (.2);
\draw (1.5,-3.2) circle (.2);

\draw (0.5,7.7) circle (.2);
\draw (1.5,7.7) circle (.2);

%

\draw (-0.3,-1.25) node (X) {$a_3$};
\draw (0.7,0.25) node (X) {$a_2$};
\draw (1.7,1.75) node (X) {$a_1$};

\draw (1.7,2.75) node (X) {$a'_1$};
\draw (0.7,4.25) node (X) {$a'_2$};
\draw (-0.3,5.75) node (X) {$a'_3$};

\draw (7,2.25) node (X) {$R^{(j)}$};
\draw[rounded corners] (6,3) rectangle (8,1.5);

\draw (5,1.75) edge[out=0,in=-90] (6.5,1.5);
\draw (4,0.25) edge[out=0,in=-90] (7,1.5);
\draw (3,-1.25) edge[out=0,in=-90] (7.5,1.5);

\draw (5,2.75) edge[out=0,in=90] (6.5,3);
\draw (4,4.25) edge[out=0,in=90] (7,3);
\draw (3,5.75) edge[out=0,in=90] (7.5,3);

\end{tikzpicture}.
\end{equation}
The algorithm for applying the quantum channel is therefore:

\begin{enumerate}
  \item Trace over the first index of $R^{(j)}$
  \begin{equation}
    R^{(j)}_{a_{1:D},a'_{1:D}}\rightarrow \sum_{a}R^{(j)}_{a a_{1:D-1},a a'_{1:D-1}}
  \end{equation}
  This reduces the number of indices of $R^{(j)}$ to $2(d-2)$, or $q^{2(d-2)}$ components.

  \item Apply the unitaries $U_{d-1}$ and $\bar U_{d-1}$. This increases the rank of the resulting tensor back to $2(d-1)$.

  \item Continue applying unitaries from the ``middle out'' for $j=d-2,\ldots 2$.

  \item Apply $U_1$ and $U^\dagger_1$, with the outer indices fixed.

\end{enumerate}

This process involves $O(d)$ steps of matrix multiplication, where the matrices are of size $O(q^d)$. Thus if one directly applies the channel to a density matrix, the overall complexity is $O(d q^{2(d-1)})$. This is better than the naive $O(d q^{3(d-1)})$ because each of the unitaries that make up the channel is a sparse matrix. A Python implementation of the algorithm is available at \url{https://github.com/AustenLamacraft/ruc}.

Since our quantum channel is constructed from a diagonal cut through the unitary circuit, there will be edge effects in a rectangular circuit of finite width. Our approach is well suited to \emph{infinite} width circuits: the channel is applied repeatedly to a random initial density matrix until a steady state density matrix is approached (for translationally invariant circuits) or a stationary distribution (for random circuits). This typically occurs on the scale of a number of steps roughly equal to the depth of the circuit.

\subsection{Low-rank approximation of $R^{(n)}$}\label{lra}

Away from fine-tuned points (see Sec.~\ref{kickedIsing}), finite-depth local unitary circuits give rise to entanglement spectra that are very broad; thus, the vast majority of the eigenstates of $\rho_A$ are close to zero and do not contribute to R\'enyi entropies with $n \geq 1$~\cite{ccgp}. This observation is implicit in the fact that different $S_n$ have different growth rates~\cite{tianci}. This fact allows us to propagate $R^{(n)}$ with negligible error using far fewer than $q^{d-1}$ basis states. We proceed as follows. We approximate

\beq\label{equur}
R^{(n)} \approx \sum_{k = 1}^K \lambda_k |k\rangle \langle k|,
\eeq
where $\lambda_k$ are the $K$ largest eigenvalues of $R$ and $|k\rangle$ are the associated eigenvectors. We renormalize all the eigenvalues to preserve the trace. We now evolve each $|k\rangle$ under each ``leg'' of the quantum channel. This evolution is efficient because each of the unitaries is a very sparse matrix. At the end of this process we have the expression

\beq\label{equus}
R^{(n+1)} \approx \sum_{k = 1}^K \sum_{i = 1}^4 \lambda_k |\phi_{ik}\rangle \langle \phi_{ik}|,
\eeq
where $|\phi_{ik}\rangle$ are not mutually orthogonal or normalized, but nevertheless span a $4K$-dimensional space. Eq.~\eqref{equus} is the ancilla density matrix that would result from one step of the quantum channel applied to the approximate density matrix~\eqref{equur}. It is a legitimate density matrix, since Eq.~\eqref{equur} was. Now we can repeat this process by approximating $R^{(n+1)}$ with its top $K$ eigenvectors, renormalizing, propagating, and so on.

When $K$ is sufficiently small, the complexity of evolving the $K$ top eigenvectors scales as $O(K d q^{d-1})$, since the unitary gates are individually sparse matrices. The diagonalization step scales as $O(K^3)$, meanwhile. Benchmarking our results against exact diagonalization at small sizes (and against exact results for random unitary circuits at arbitrary sizes) we find that keeping $\alt 100$ eigenvectors suffices to capture the quantities that are of interest here---mainly, the purity and min-entropy, and their fluctuations.

\subsection{Trajectory approach for computing the purity}\label{sec:traj}

In this section we describe an approach based on mapping \emph{vectors} in the ancilla space rather than density matrices. Formally, this method is equivalent to the trajectory approach developed to analyze master equations in quantum optics~\cite{carmichael_book}. Applying unitaries to a vector is an $O(d q^{(d-1)})$ operation because the unitaries are sparse. While such an approach is evidently attractive, we will see that there is a trade-off in terms of the number of times the matrices $A^{(j)}_{s_1,s_2}$ must be applied.

The ancilla density matrix can be defined in terms of averages over trajectories in the physical indices in the following way. Starting from the Kraus form
\begin{equation}\label{eq:kraus}
R^{(j)} = \sum_s A^{(j)\vphantom{\dagger}}_{s} R^{(j-1)} A^{(j)\dagger}_{s}, \qquad j=n+1,\ldots N,
\end{equation}
In the case of unitary circuits the index $s$ is a composite: $s=(s_1,s_2)$, and we have changed the indexing of slices so that indices increase going right to left. We see that $L$ updates correspond to summing over trajectories in the physical indices of length $L$
\begin{equation}
  R^{(L)} = \sum_{s_1:s_L\in \{\mathbb{Z}_q\}^L} A^{(L)\vphantom{\dagger}}_{s_L}\cdots  A^{(1)\vphantom{\dagger}}_{s_1} R^{(0)} A^{(1)\dagger}_{s_1}\cdots A^{(L)\dagger}_{s_L}.
\end{equation}
If we start from a pure state $R^{(0)}=\ket{\psi_0}\bra{\psi_0}$ we can write this as an average over trajectories with uniform distribution
\begin{equation}\label{eq:exp}
  R^{(L)} = q^L \E_{s_{1:L}\sim \text{uniform}}\left[\ket{\tilde\psi_{s_{1:L}}}\bra{\tilde\psi_{s_{1:L}}}\right],
\end{equation}
where the vectors $\ket{\psi_{s_{1:L}}}$ are defined as
\begin{equation}
  \ket{\tilde \psi_{s_{1:L}}} = A^{(L)}_{s_L}\cdots A^{(1)}_{s_1}\ket{\psi_0}.
\end{equation}
These vectors are unnormalized. The normalization factors
\begin{equation}
  p(s_1:s_L)\equiv\braket{\tilde\psi_{s_{1:L}}|\tilde\psi_{s_{1:L}}}
\end{equation}
are a normalized probability distribution over trajectories by virtue of the left canonical condition Eq.~\eqref{eq:left-c}. Denoting the normalized vectors as $\ket{\psi_{s_{1:L}}}$ we can express the ancilla density matrix as
\begin{equation}\label{eq:exp_traj}
  R^{(L)}=\E_{s_{1:N}\sim p(\cdot) }\left[\ket{\psi_{s_{1:L}}}\bra{\psi_{s_{1:L}}}\right].
\end{equation}
As a trajectory increases in length, the normalization factors are updated according to
\begin{equation}
  p(s_1:s_L) = p(s_1:s_{L-1})\braket{\psi_{L-1}|A^{(L)\dagger}_{s_L}A^{(L)}_{s_L}|\psi_{L-1}},
\end{equation}
so that the second factor may be interpreted as a conditional probability
\begin{equation}\label{eq:trans}
  p(s_L|s_{1:L-1})=\braket{\psi_{L-1}|A^{(L)\dagger}_{s_L}A^{(L)}_{s_L}|\psi_{L-1}},
\end{equation}
Eq.~\eqref{eq:exp_traj} expresses the ancilla density matrix in terms of vectors, but it requires an average over trajectories. The downside of this approach is that evaluating $R$ will require roughly $\gamma^{-1}$ trajectories, where $\gamma$ is the purity of a half-infinite region, which sets the approximate rank of the reduced density matrix. However, this approach lends itself to parallelization while the channel based approach does not.

For a translationally invariant system, and assuming this random process is ergodic, we can substitute an average over the length of single long trajectory in the $L\to\infty$ limit.
\begin{equation}\label{eq:ergodic}
  R = \lim_{L\to\infty} \frac{1}{L}\sum_{l=1}^L \ket{\psi_{s_{1:l}}}\bra{\psi_{s_{1:l}}}.
\end{equation}
In practice, long runs and multiple trajectories are used.

$O(D)$ evaluation of matrix products is not much use if we still need $O(D^3)$ evaluation of the spectrum of $R^{(L)}$ or $O(D^2)$ evaluation of the purity.  However, we can access the purity without dealing with $R^{(L)}$ directly using
\begin{equation}\label{eq:pure_traj}
  \gamma_L=\tr[R^{(L)2}]=\E_{s_{1:N},t_{1:N}\sim p(\cdot)}|\braket{\psi_{t_{1:L}}|\psi_{s_{1:L}}}|^2,
\end{equation}
which follows from Eq.~\eqref{eq:exp_traj}. This formula expresses the purity as the average fidelity over pairs of trajectories. In a high purity state the ancilla vectors stay close to each other as they evolve over different trajectories, whereas in a highly entangled state different trajectories explore different regions of ancilla space.

Note that the expectations discussed in this section are unrelated to any random variables that may form part of the specification of the circuit. Evaluating the average purity, for example, would require an additional average of Eq.~\eqref{eq:pure_traj} over these variables.


\section{Exactly solvable example: self-dual kicked Ising model}\label{kickedIsing}

In order to illustrate the utility of the formalism introduced in the previous section, we now turn to an example of a unitary circuit in which the entanglement spectrum can be determined analytically. This is the kicked Ising model at the self-dual point discussed in two recent papers \cite{Bertini:2018aa,Bertini:2018fbz}. The kicked Ising model describes the evolution of a system of $L$ spin-1/2 subsystems (qubits) for an integer time $t$ by the unitary operator $\left(U_\text{KI}\right)^t$, where $U_\text{KI}=K I_\mathbf{h}$ is composed of the two unitaries
\begin{equation}\label{eq:KIM}
  I_\mathbf{h} = e^{-iH_\text{I}[\mathbf{h}]},\qquad
  K = e^{-iH_\text{K}},
\end{equation}
where
\begin{align}
H_\text{I}[\mathbf{h}]&=\sum_{j=1}^L\left[J Z_j Z_{j+1} + h_j Z_j\right]\\
H_\text{K} &= b\sum_{j=1}^L X_j,
\end{align}
and $(X_j,Y_j,Z_j)$ are the Pauli matrices for spin $j$. $H_\text{I}[\mathbf{h}]$ is the classical Ising model with arbitrary longitudinal fields $h_j$, while $H_\text{K}$ describes a transverse field.

In Ref.~\cite{Bertini:2018fbz} the growth of the entanglement entropies was found exactly for some particular initial product states at the special `self-dual' values
\begin{equation}\label{eq:self_dual}
  |J|=|b|=\frac{\pi}{4}.
\end{equation}
For a region $A$ of size $N$, the authors found that when starting from an arbitrary product state in the $Z_j$ basis the R\'enyi entropies \eqref{eq:ren_def} to be \emph{exactly} given by
\begin{equation}\label{eq:bertini_res}
  \lim_{L\to\infty} S^{(n)}_A(t) =\min(2t-2,N)\log 2,
\end{equation}
\emph{independent} of R\'enyi index $n$. The interpretation in terms of the entanglement spectrum is striking: there are $2^{\min(2t-2,N)}$ eigenvalues equal to $2^{-\min(2t-2,N)}$ and the rest are zero.

We now show how our quantum channel approach may be used to derive the corresponding result for the case of a semi-infinite interval
\begin{equation}\label{eq:our_res}
  \lim_{L\to\infty} S^{(n)}_A(t) =(t-1)\log 2.
\end{equation}
We can present the unitary $(U_\text{KI})^t$ as a unitary circuit of depth $t$, where the layers alternate between unitaries operating between spins $2j-1$ and $2j$, and between spins $2j$ and $2j+1$. A variety of decompositions are available. For reasons that will become clear, we choose the following:
\begin{equation}\begin{tikzpicture}[baseline={([yshift=-1ex]current bounding box.center)},every node/.style={scale=1},scale=.75]
  \mpsT{0}{0}{U_{12}}
\end{tikzpicture}=
\begin{tikzpicture}[baseline={([yshift=-1ex]current bounding box.center)},every node/.style={scale=1},scale=.75]

  \draw[rounded corners] (0,2) rectangle (3,1);
  \draw (1.5,1.5) node (X) {$\cI$};

  \draw[rounded corners] (0,-2) rectangle (3,-1);
  \draw (1.5,-1.5) node (X) {$\cI$};

  \mpsT{0.5}{0}{\cK}
  \mpsT{2.5}{0}{\cK}

  \lineV{1}{0.5}{0.5}
  \lineV{-1}{-0.5}{0.5}
  \lineV{1}{0.5}{2.5}
  \lineV{-1}{-0.5}{2.5}

  \lineV{2}{2.5}{0.5}
  \lineV{2}{2.5}{2.5}
  \lineV{-2}{-2.5}{0.5}
  \lineV{-2}{-2.5}{2.5}

  \draw (0.5,2.8) node (X) {$a$};
  \draw (2.5,2.8) node (X) {$b$};

  \draw (0.5,-2.8) node (X) {$c$};
  \draw (2.5,-2.8) node (X) {$d$};

\end{tikzpicture},
\end{equation}
where the one qubit ($\cK$) and two qubit ($\cI$) gates have the form
\begin{align}
  \cK &= \exp\left[-i b X\right]\\
  \cI &= \exp\left[-iJ Z_1 Z_2 -i \left(h_1 Z_1 + h_2 Z_2\right)/2\right].
\end{align}
In the $Z$ basis, the elements of $U_{12}$ are
\begin{multline}\label{eq:elem}
  (U_{12})_{ab,cd} =-\frac{\sin 2b}{2} \exp\left(-iJ [ab+cd]-i\tilde J[ac+bd]\right)\\
  \qquad\times \exp\left(-ih_1[a+c]/2-ih_2[b+d]/2\right)
\end{multline}
where $a,b,c,d\in\{1,-1\}$ and
\begin{equation}\label{eq:dualJ}
  \tilde J = -\frac{\pi}{4}-\frac{i}{2}\log\tan b.
\end{equation}
%
%
The matrix $(U_{12})_{ab,cd}$ is unitary, but the matrix $\tilde U_{12}$ with elements $(\tilde U)_{ab,cd}=(U_{12})_{ac,bd}$ is not \emph{except} at the self-dual points Eq.~\eqref{eq:self_dual}. The unitarity of $\tilde U_{12}$ has the graphical representation
\begin{equation}\label{eq:utilde}
  \begin{tikzpicture}[baseline={([yshift=-1ex]current bounding box.center)},every node/.style={scale=1},scale=.55]

  \mpsT{1}{1.5}{U_{12}}

  \mpsT{1}{-1.5}{U_{12}^\dagger}

  \draw (1.5,2) edge[out=90,in=90] (2.5,2);
  \draw (2.5,2) edge[out=-90,in=90] (2.5,-2);
  \draw (1.5,-2) edge[out=-90,in=-90] (2.5,-2);


  \lineV{2}{-2}{1.5}

  \draw (0,2.5) edge[out=0,in=90] (0.5,2);
  \draw (0,0.5) edge[out=0,in=-90] (0.5,1);

  \draw (0,-2.5) edge[out=0,in=-90] (0.5,-2);
  \draw (0,-0.5) edge[out=0,in=90] (0.5,-1);

  \lineV{1}{2}{0.5}
  \lineV{-1}{-2}{0.5}

  \draw (-0.3,2.5) node (X) {$a$};
  \draw (-0.3,0.5) node (X) {$b$};

  \draw (-0.3,-2.5) node (X) {$a'$};
  \draw (-0.3,-0.5) node (X) {$b'$};
  \end{tikzpicture}=
  \begin{tikzpicture}[baseline={([yshift=-1ex]current bounding box.center)},every node/.style={scale=1},scale=.55]

  \draw (0,2.5) edge[out=0,in=90] (1,2);
  \draw (0,0.5) edge[out=0,in=90] (0.5,0);

  \draw (0,-2.5) edge[out=0,in=-90] (1,-2);
  \draw (0,-0.5) edge[out=0,in=-90] (0.5,0);

  \lineV{2}{-2}{1}

  \draw (-0.3,2.5) node (X) {$a$};
  \draw (-0.3,0.5) node (X) {$b$};

  \draw (-0.3,-2.5) node (X) {$a'$};
  \draw (-0.3,-0.5) node (X) {$b'$};
  \end{tikzpicture}
  = \delta_{aa'}\delta_{bb'}.
  \end{equation}
The unitarity of $\tilde U_{12}$ has the interesting consequence that the SW-NE MPS is in left \emph{and right} canonical form, so that
\begin{equation}\label{eq:unital}
  \sum_s A^{(j)\vphantom{\dagger}}_{s}A^{(j)\dagger}_{s} = \openone_{2^{t-1}}.
\end{equation}
To see this, we first give the graphical representation of the left hand side of Eq.~\eqref{eq:unital}
\begin{widetext}
\begin{equation}\sum_{s_1,s_2} \left(A^{(j)}_{s_1,s_2}A^{(j)^\dagger}_{s_1,s_2}\right)_{a_{1:3},a'_{1:3}}=
\begin{tikzpicture}[baseline={([yshift=-1ex]current bounding box.center)},every node/.style={scale=1},scale=.55]

\mpsT{1}{1.5}{U_{12}}
\mpsT{2}{3}{U_{23}}
\mpsT{3}{4.5}{U_{34}}
\mpsT{4}{6}{U_{45}}

\mpsT{1}{-1.5}{U_{12}^\dagger}
\mpsT{2}{-3}{U_{23}^\dagger}
\mpsT{3}{-4.5}{U_{34}^\dagger}
\mpsT{4}{-6}{U_{45}^\dagger}

\draw (3.5,6.5) edge[out=90,in=90] (6.5,6.5);
\draw (6.5,6.5) edge[out=-90,in=90] (6.5,-6.5);
\draw (3.5,-6.5) edge[out=-90,in=-90] (6.5,-6.5);

\draw (4.5,6.5) edge[out=90,in=90] (5.5,6.5);
\draw (5.5,6.5) edge[out=-90,in=90] (5.5,-6.5);
\draw (4.5,-6.5) edge[out=-90,in=-90] (5.5,-6.5);

\lineV{6.5}{-6.5}{4.5}
\lineV{6.5}{-6.5}{3.5}
\lineV{5}{-5}{2.5}
\lineV{3.5}{0.5}{1.5}
\lineV{-3.5}{-0.5}{1.5}
\lineV{2}{0.5}{0.5}
\lineV{-2}{-0.5}{0.5}

\draw (0.5,0.3) circle (.2);
\draw (1.5,0.3) circle (.2);
\draw (0.5,-0.3) circle (.2);
\draw (1.5,-0.3) circle (.2);

\draw (0,2.5) edge[out=0,in=90] (0.5,2);
\draw (1,4) edge[out=0,in=90] (1.5,3.5);
\draw (2,5.5) edge[out=0,in=90] (2.5,5);

\draw (0,-2.5) edge[out=0,in=-90] (0.5,-2);
\draw (1,-4) edge[out=0,in=-90] (1.5,-3.5);
\draw (2,-5.5) edge[out=0,in=-90] (2.5,-5);

\draw (-0.3,2.5) node (X) {$a_1$};
\draw (0.7,4) node (X) {$a_2$};
\draw (1.7,5.5) node (X) {$a_3$};

\draw (-0.3,-2.5) node (X) {$a_1'$};
\draw (0.7,-4) node (X) {$a_2'$};
\draw (1.7,-5.5) node (X) {$a_3'$};

\end{tikzpicture}=
\begin{tikzpicture}[baseline={([yshift=-1ex]current bounding box.center)},every node/.style={scale=1},scale=.55]

\mpsT{1}{1.5}{U_{12}}
\mpsT{2}{3}{U_{23}}
\mpsT{3}{4.5}{U_{34}}

\mpsT{1}{-1.5}{U_{12}^\dagger}
\mpsT{2}{-3}{U_{23}^\dagger}
\mpsT{3}{-4.5}{U_{34}^\dagger}

\draw (3.5,5) edge[out=90,in=90] (4.5,5);
\draw (4.5,5) edge[out=-90,in=90] (4.5,-5);
\draw (3.5,-5) edge[out=-90,in=-90] (4.5,-5);


\lineV{5}{-5}{3.5}
\lineV{5}{-5}{2.5}
\lineV{3.5}{0.5}{1.5}
\lineV{-3.5}{-0.5}{1.5}
\lineV{2}{0.5}{0.5}
\lineV{-2}{-0.5}{0.5}

\draw (0.5,0.3) circle (.2);
\draw (1.5,0.3) circle (.2);
\draw (0.5,-0.3) circle (.2);
\draw (1.5,-0.3) circle (.2);

\draw (0,2.5) edge[out=0,in=90] (0.5,2);
\draw (1,4) edge[out=0,in=90] (1.5,3.5);
\draw (2,5.5) edge[out=0,in=90] (2.5,5);

\draw (0,-2.5) edge[out=0,in=-90] (0.5,-2);
\draw (1,-4) edge[out=0,in=-90] (1.5,-3.5);
\draw (2,-5.5) edge[out=0,in=-90] (2.5,-5);

\draw (-0.3,2.5) node (X) {$a_1$};
\draw (0.7,4) node (X) {$a_2$};
\draw (1.7,5.5) node (X) {$a_3$};

\draw (-0.3,-2.5) node (X) {$a_1'$};
\draw (0.7,-4) node (X) {$a_2'$};
\draw (1.7,-5.5) node (X) {$a_3'$};

\end{tikzpicture}
\end{equation}
\end{widetext}
Where we used unitarity to eliminate the top and bottom gates. Using the motif Eq.~\eqref{eq:utilde} we can telescope telescope the circuit until
\begin{equation}\label{frag58}
  \sum_{s_1,s_2} \left(A^{(j)}_{s_1,s_2}A^{(j)^\dagger}_{s_1,s_2}\right)_{a_{1:3},a'_{1:3}}=
\begin{tikzpicture}[baseline={([yshift=-1ex]current bounding box.center)},every node/.style={scale=1},scale=.55]

\mpsT{1}{1.5}{U_{12}}

\mpsT{1}{-1.5}{U_{12}^\dagger}

\draw (1.5,2) edge[out=90,in=90] (2.5,2);
\draw (2.5,2) edge[out=-90,in=90] (2.5,-2);
\draw (1.5,-2) edge[out=-90,in=-90] (2.5,-2);

\lineV{5}{-5}{3.5}
\lineV{3.55}{-3.5}{3}

\lineV{2}{0.5}{1.5}
\lineV{-2}{-0.5}{1.5}

\lineV{2}{0.5}{0.5}
\lineV{-2}{-0.5}{0.5}

\draw (0.5,0.3) circle (.2);
\draw (1.5,0.3) circle (.2);
\draw (0.5,-0.3) circle (.2);
\draw (1.5,-0.3) circle (.2);

\draw (0,2.5) edge[out=0,in=90] (0.5,2);

\draw (1,4) edge[out=0,in=90] (3,3.5);
\draw (2,5.5) edge[out=0,in=90] (3.5,5);

\draw (0,-2.5) edge[out=0,in=-90] (0.5,-2);

\draw (1,-4) edge[out=0,in=-90] (3,-3.5);
\draw (2,-5.5) edge[out=0,in=-90] (3.5,-5);

\draw (-0.3,2.5) node (X) {$a_1$};
\draw (0.7,4) node (X) {$a_2$};
\draw (1.7,5.5) node (X) {$a_3$};

\draw (-0.3,-2.5) node (X) {$a_1'$};
\draw (0.7,-4) node (X) {$a_2'$};
\draw (1.7,-5.5) node (X) {$a_3'$};

\end{tikzpicture}
\end{equation}
Finally, we use the explicit form of Eq.~\eqref{eq:elem} at the self-dual point to evaluate
\begin{equation}
  \begin{tikzpicture}[baseline={([yshift=-1ex]current bounding box.center)},every node/.style={scale=1},scale=.55]

  \mpsT{1}{1.5}{U_{12}}

  \mpsT{1}{-1.5}{U_{12}^\dagger}

  \draw (1.5,2) edge[out=90,in=90] (2.5,2);
  \draw (2.5,2) edge[out=-90,in=90] (2.5,-2);
  \draw (1.5,-2) edge[out=-90,in=-90] (2.5,-2);

  \lineV{2}{0.75}{1.5}
  \lineV{-2}{-0.75}{1.5}

  \lineV{2}{0.75}{0.5}
  \lineV{-2}{-0.75}{0.5}

  \draw (0.5,0.55) circle (.2);
  \draw (1.5,0.55) circle (.2);
  \draw (0.5,-0.55) circle (.2);
  \draw (1.5,-0.55) circle (.2);

  \draw (0,2.5) edge[out=0,in=90] (0.5,2);
  \draw (0,-2.5) edge[out=0,in=-90] (0.5,-2);

  \draw (-0.3,2.5) node (X) {$a_1$};
  \draw (-0.3,-2.5) node (X) {$a_1'$};

\draw (0.5,0) node (X) {$c$};
\draw (1.5,0) node (X) {$d$};

  \end{tikzpicture}=
  \sum_{b} \left(U_{12}\right)_{ab,cd} \left(U^*_{12}\right)_{a'b,cd}=\delta_{aa'},
\end{equation}
which corresponds to an initial state equal to a product state in the $Z_j$ basis. This verifies the condition Eq.~\eqref{eq:unital}.

An MPS in both left and right canonical form describes a \emph{bistochastic} quantum channel: one that preserves the identity. As a result, the ancilla density matrix $R^{j}=2^{1-t}\openone_{2^{t-1}}$ for all $j$. Evaluating the R\'enyi entropies yields our result Eq.~\eqref{eq:our_res}.


It seems likely that a similar analysis can be performed for general Clifford circuits, which also have degenerate entanglement spectra~\cite{nrvh}. However, establishing this in general requires one to carve out several special cases (such as circuits that generate no entanglement at all from a number of initial states~\cite{gz2018, sg_og_2018}) and we will not pursue this here.

\section{Numerical results: random unitary circuits}\label{results}

In this section we present numerical results for the evolution of various entanglement measures, and their spatial fluctuations, for random unitary circuits. The coarse features of the evolution of the entanglement spectrum were already discussed in Ref.~\cite{ccgp}. In particular, the bandwidth of the entanglement spectrum broadens linearly in time; as noted in that work, this broadening is a natural consequence of the wide separation between the entanglement and light-cone speeds. For circuits of depth $\geq 10$ this broadening implies that an appreciable fraction of the spectrum of the reduced density matrix is zero to within numerical precision. Our focus here is on the \emph{large} eigenvalues of the reduced density matrix (which dominate R\'enyi entropies $S_n, n \geq 1$). Because the vast majority of the eigenvalues are near zero, this ``low-entanglement-energy'' tail can be described accurately by low-rank approximations as in Sec.~\ref{lra}, allowing us to go to circuits of depth $L = 14$ with minor computational effort.

\subsection{Benchmarking the low-rank approximation}

\begin{figure}[tb]
\begin{center}
\includegraphics[width=0.4\textwidth]{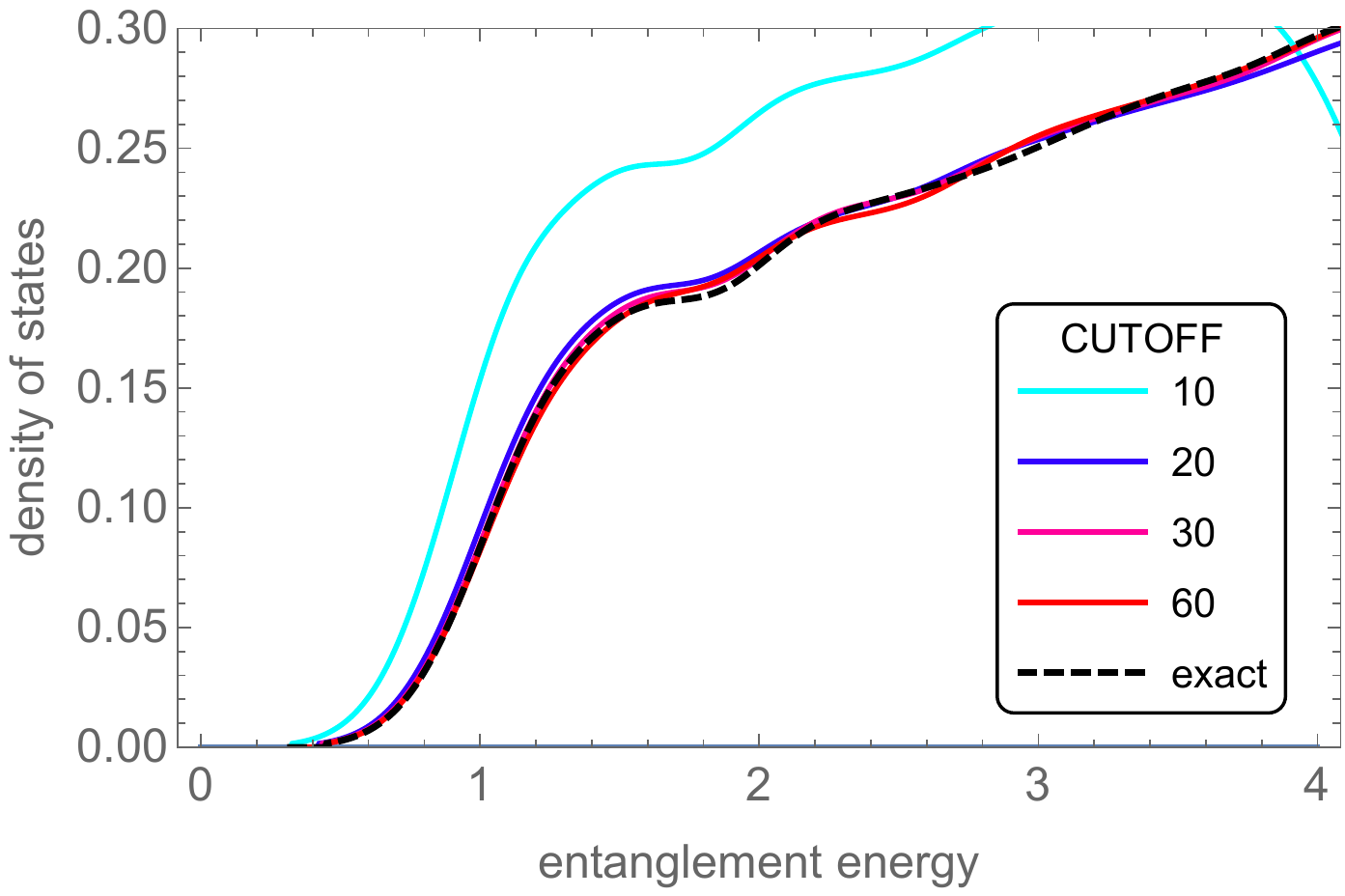}
\caption{Entanglement density of states (i.e., histogram of entanglement energies), at depth $t = 8$, computed both exactly and by low-rank approximation of the quantum channel (Sec.~\ref{lra}). For rank $\agt 20$ the low-energy behavior of the entanglement spectrum is well captured.}
\label{vscutoff}
\end{center}
\end{figure}

For depths $t \leq 10$ we can compare the entanglement spectra computed by low-rank approximation with the exact ones (Fig.~\ref{vscutoff}). Although the rank of the reduced density matrix in this case is $256$, we find that working with the top 20 states allows us to match the low-energy behavior of the entanglement spectrum.

For the largest depths we have considered, exact time evolution is not feasible; however, there is an exact result for the average purity of a semi-infinte system~\cite{Nahum2017}, viz. $\bar\gamma = (4/5)^{t-1}$ \footnote{The exponent $t-1$ here is one less than in \cite{Nahum2017} because our partition of the system lies between the two gates in the top layer of the circuit, which therefore leaves the entanglement unaffected.}. For the numerical results presented here we increase the rank of the approximation until the mean computed purity matches this exact result to within statistical error (which is about $1\%$). For the largest depth we have systematically considered ($t = 14$) we need to keep $\approx 120$ states to match the mean purity. This is only about $1\%$ of the spectrum; thus the low-rank approximation is much more efficient than direct propagation of the channel would be.

\subsection{Shape of the entanglement spectrum}

In this section we discuss the shape of the entanglement spectrum and its relation to the evolution of the R\'enyi entropies for RUCs. First, let us recall that the eigenstate thermalization hypothesis predicts that the reduced density matrix of a subsystem should take the form $\mathcal{N} \exp(- \beta H)$, where $H$ is the Hamiltonian of the subsystem and $\beta$ is the inverse temperature. Therefore the entanglement Hamiltonian $-\log \rho \propto \beta H + \log \mathcal{N}$, i.e., it is a stretched and shifted version of the physical Hamiltonian. The entanglement spectrum therefore has the same shape as the physical spectrum: for a large subsystem $L_A$ with a local Hamiltonian, it will be essentially Gaussian in the bulk, with a bandwidth that increases as $\sqrt{L_A}$, although the extreme value statistics (corresponding to the shape of the spectrum near its ground state) are model-dependent. When $\beta \sqrt{L_A}$ is large, the entanglement spectrum will have a large bandwidth, and therefore (because of the Jacobian) the reduced density matrix will have a density of eigenvalues $\varrho(\lambda)$ distribution of the form $\varrho(\lambda) \sim 1/\lambda$. Infinite temperature is a singular limit, as $\beta = 0$ so the entanglement spectrum is degenerate. In practice, a typical, randomly picked state deviates form infinite temperature by an amount $\sim 1/\sqrt{L_A}$, so the entanglement ``bandwidth'' is $L_A$-independent (up to possible logarithmic dependences that we are not concerned with here). For Floquet systems or RUCs with no conservation laws, these arguments suggest that the entanglement spectrum of a small subsystem at very late times is degenerate up to finite size effects. Once finite size effects are included we expect a Marchenko--Pastur distribution~\cite{Chen:2017aa}.

\begin{figure}[!t!b]
\begin{center}
\includegraphics[width = 0.45\textwidth]{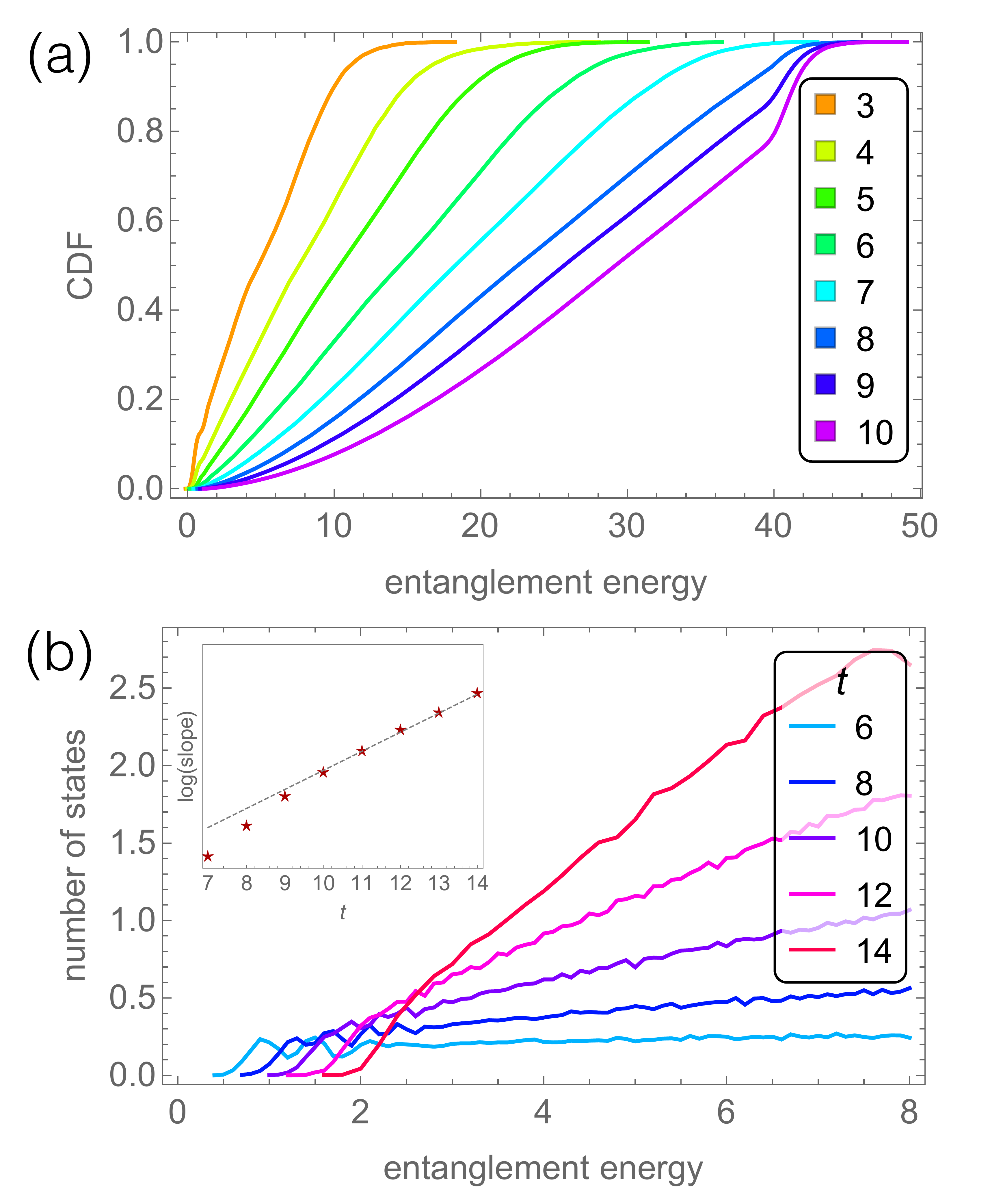}
\caption{(a)~Cumulative distribution function of the entanglement eigenvalues under random unitary dynamics, for various circuit depths, extracted from exact evolution of the reduced density matrix under the quantum channel corresponding to the unitary circuit. At the largest depths, an appreciable fraction of the entanglement spectrum consists of eigenvalues below machine precision. (b)~A more detailed view of the entanglement spectrum near its ``low-energy'' edge at various depths $t$ under random unitary dynamics. Recall that the rank of the density matrix increases exponentially with the depth. All points except $t = 14$ are averaged over $3000$ realizations; $t = 14$ is averaged over $300$ realizations. Inset: slope of the linear growth of the entanglement density of states with energy.}
\label{entedge}
\end{center}
\end{figure}

For the case of interest to us -- large subsystems at short times -- the numerical evidence~\cite{ccgp} suggests that the spectrum of the reduced density matrix has the density $\varrho(\lambda) \sim 1/\lambda$ over many decades, at any time $1 \ll t \ll l_A$; this is qualitatively unlike the (compact) Wishart distribution that obtains at very late times~\cite{Chen:2017aa}. The bulk of the spectrum of the reduced density matrix consists of eigenvalues below machine precision whenever $t \agt 8$ (Fig.~\ref{entedge}). Here, we are concerned with the ``low-energy'' or ``high-Schmidt-coefficient'' edge, which governs the behavior of the R\'enyi entropies $S^{(n)}, n \geq 1$, which we can follow out to later times $t \approx 14$ (Fig.~\ref{entedge}). The evolution at early times is nontrivial, but appears to settle down into a well-defined limiting behavior for $t \agt 7$: there is a threshold in the entanglement density of states, followed by a linear increase with entanglement energy that persists out to the energies we can reliably access. The coefficient of this linear growth is approximately $2^{t/2}$ for the larger accessible $t$. This is exponentially slower than the growth of the total number of states, so the fraction of states in the tails thins out exponentially in time.

\begin{figure}[tb]
\begin{center}
\includegraphics[width = 0.45\textwidth]{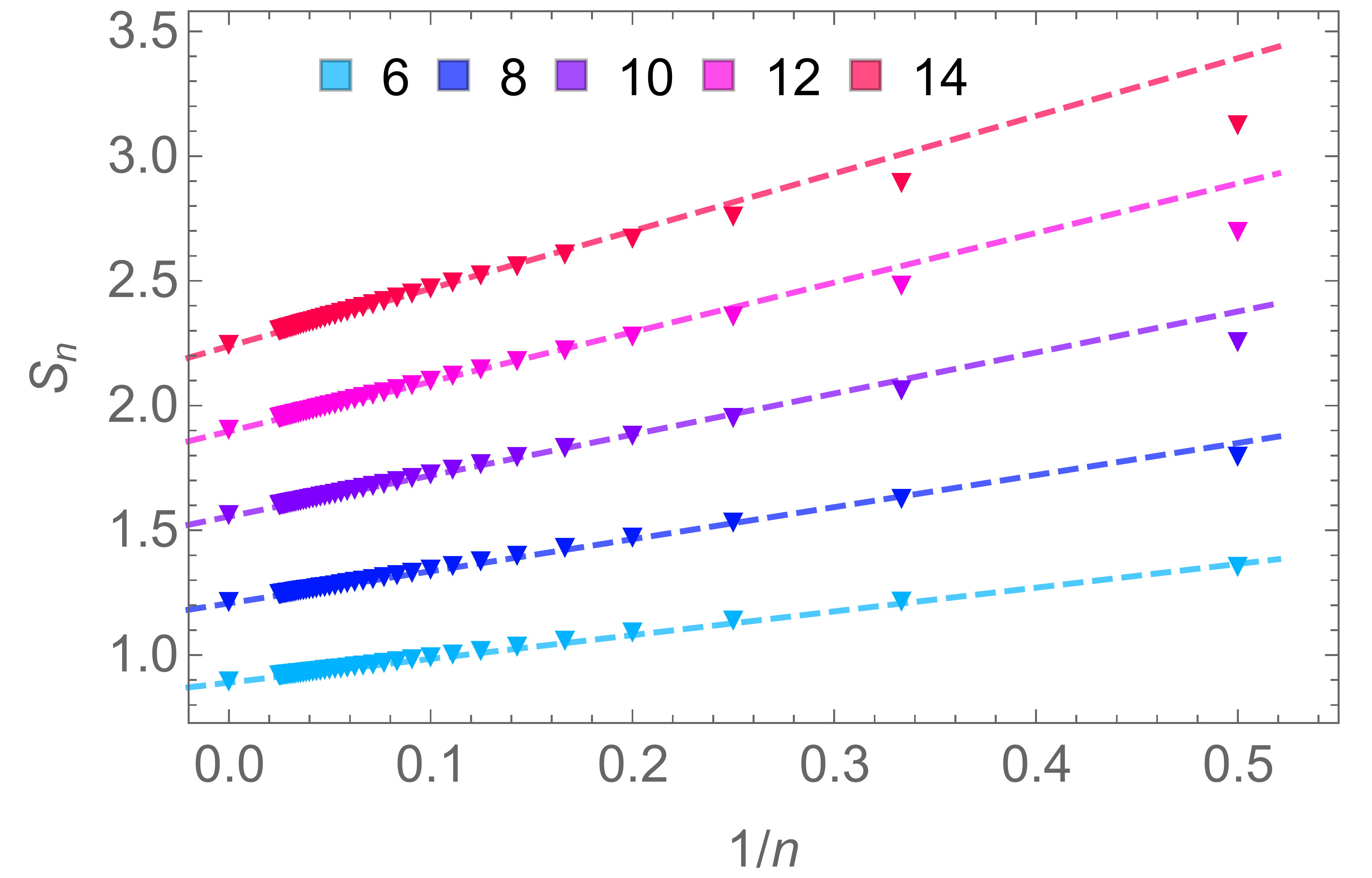}
\caption{R\'enyi-index-dependence of the entropy $S_\alpha$. The fits are to the form $S_n = S_\infty (1 - 1/n)$, which works for large $n$, with $S_\infty \simeq t/6$. Deviations from this form grow with time.}
\label{renyidep}
\end{center}
\end{figure}


The R\'enyi index dependence of $S^{(n)}$ for large $n$ follows from this behavior of the limit shape, if we further assume that the entanglement spectrum is self-averaging.
A simple model for the spectral density $\rho(\epsilon)$ of the entanglement energies that is consistent with the large deviation form Eq.~\eqref{eq:large_dev} is
\begin{equation}
  \rho(\epsilon) = \exp\left[\alpha t\Theta(E-v_\infty t)\right],
\end{equation}
in which case
\begin{equation}
  v_n = \frac{\alpha - nv_\infty}{1-n}
\end{equation}
A large $n$ result this implies $v_n/v_\infty \simeq 1 - 1/n$, which is consistent with our numerical observations [Fig.~\ref{renyidep}].


Although the late-time entanglement spectrum is not numerically accessible, our results allow us to comment on a few possible qualitative scenarios of the entanglement spectrum. First, it is clear numerically that the probability density of states is exponentially small near the low-energy edge of the entanglement spectrum. This turns out to be necessary for the R\'enyi entropies to have distinct velocities. (If one considers, e.g., a box-shaped entanglement DOS, it is simple to show that all R\'enyi entropies with $n > 0$ must have the same velocity, regardless of the aspect ratio of the box. Similar results hold for Marchenko-Pastur and other possible compact shapes.) Second, one might suppose the entanglement spectrum has a Gaussian shape. Matching exact results for $S_0$, the normalization of $\rho$, and $S_2$ requires the Gaussian to have a linearly growing mean and variance. Numerically, the entanglement bandwidth grows linearly in time rather than as a square root, possibly because of level repulsion. Developing a theory of how the entanglement spectrum evolves is an important question for future work: at present we do not have even a phenomenological Brownian-motion model of this growth.

\subsection{Statistics and spatial correlations of entanglement}

\begin{figure}[tb]
\begin{center}
\includegraphics[width = 0.45\textwidth]{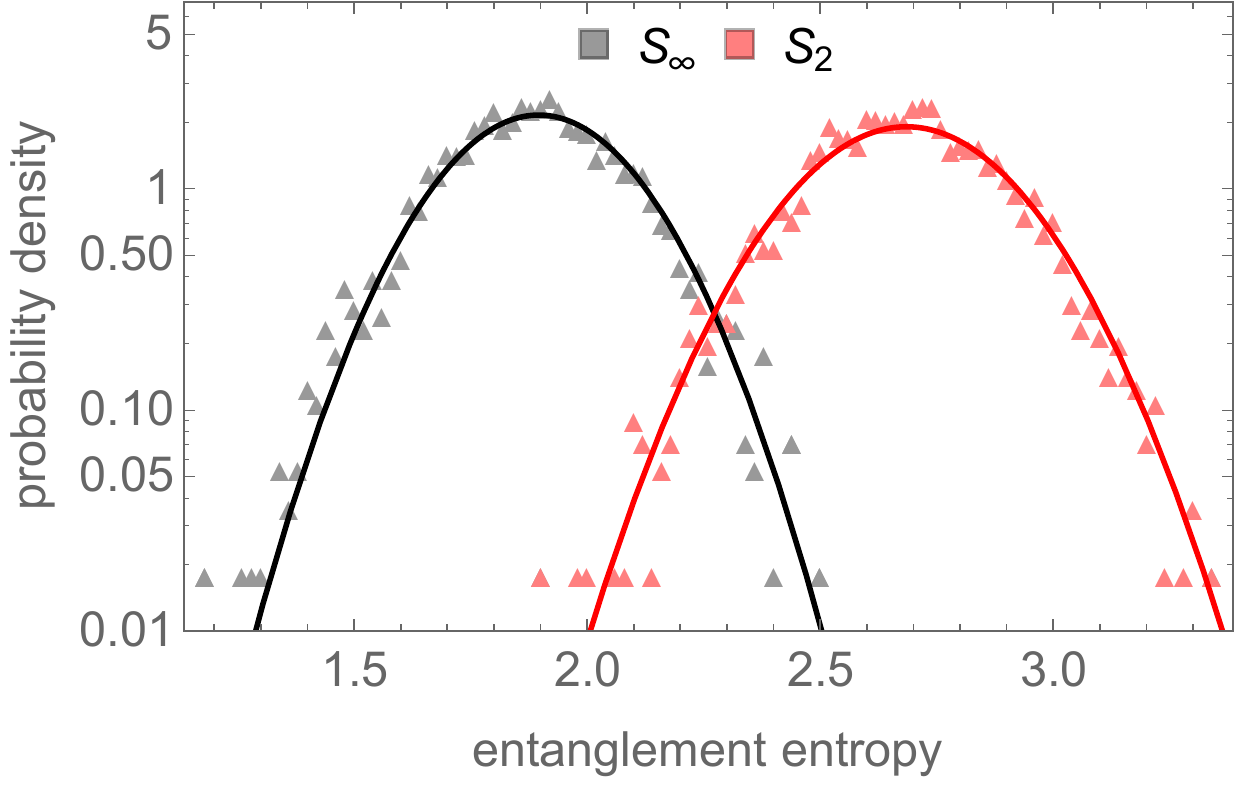}
\caption{Histograms of $S_2$ and $S_\infty$ for $t = 12$; lines are fits to a Gaussian.}
\label{enthist}
\end{center}
\end{figure}

Out to the latest times we have considered, the sample-averaged R\'enyi entropies have Gaussian distributions. (Thus, quantities such as the purity are log-normally distributed.) Whether these distributions become anisotropic at much larger system sizes is unclear; however, we have not seen any sign of incipient skewness out to the times we can simulate (Fig.~\ref{enthist}).

\begin{figure}[tb]
\begin{center}
\includegraphics[width=0.4\textwidth]{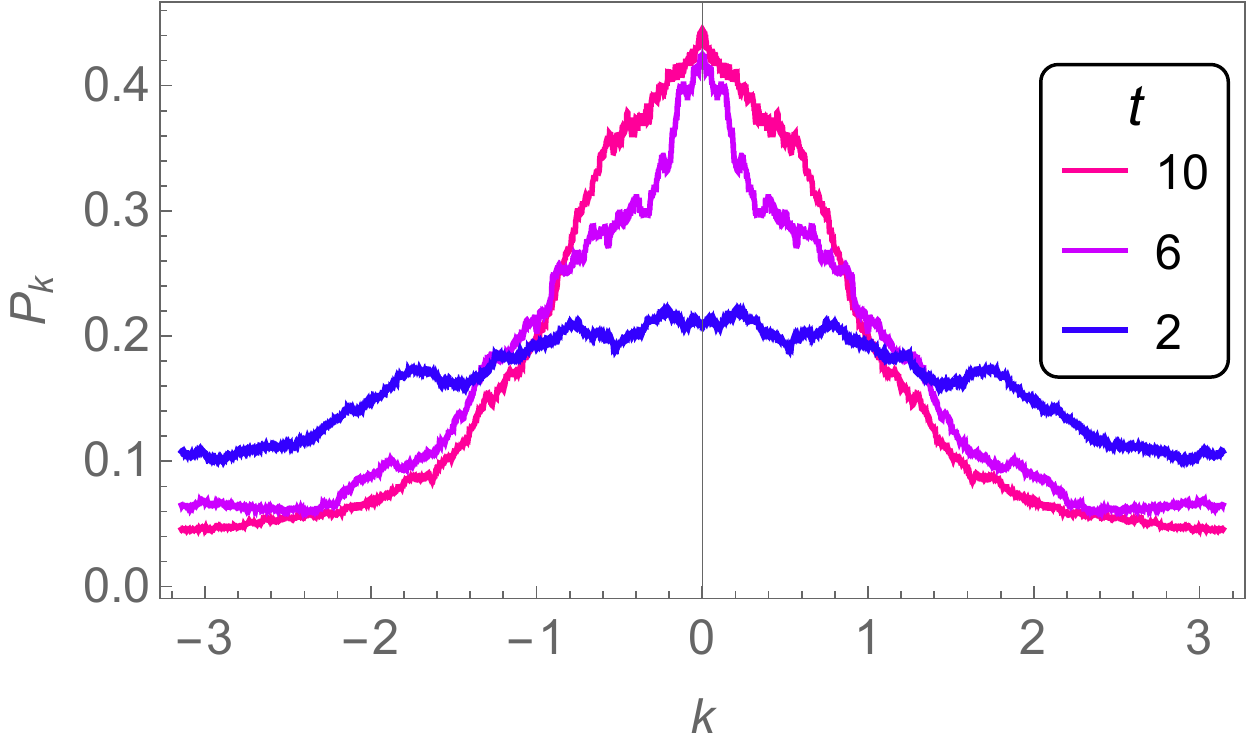}
\includegraphics[width = 0.4\textwidth]{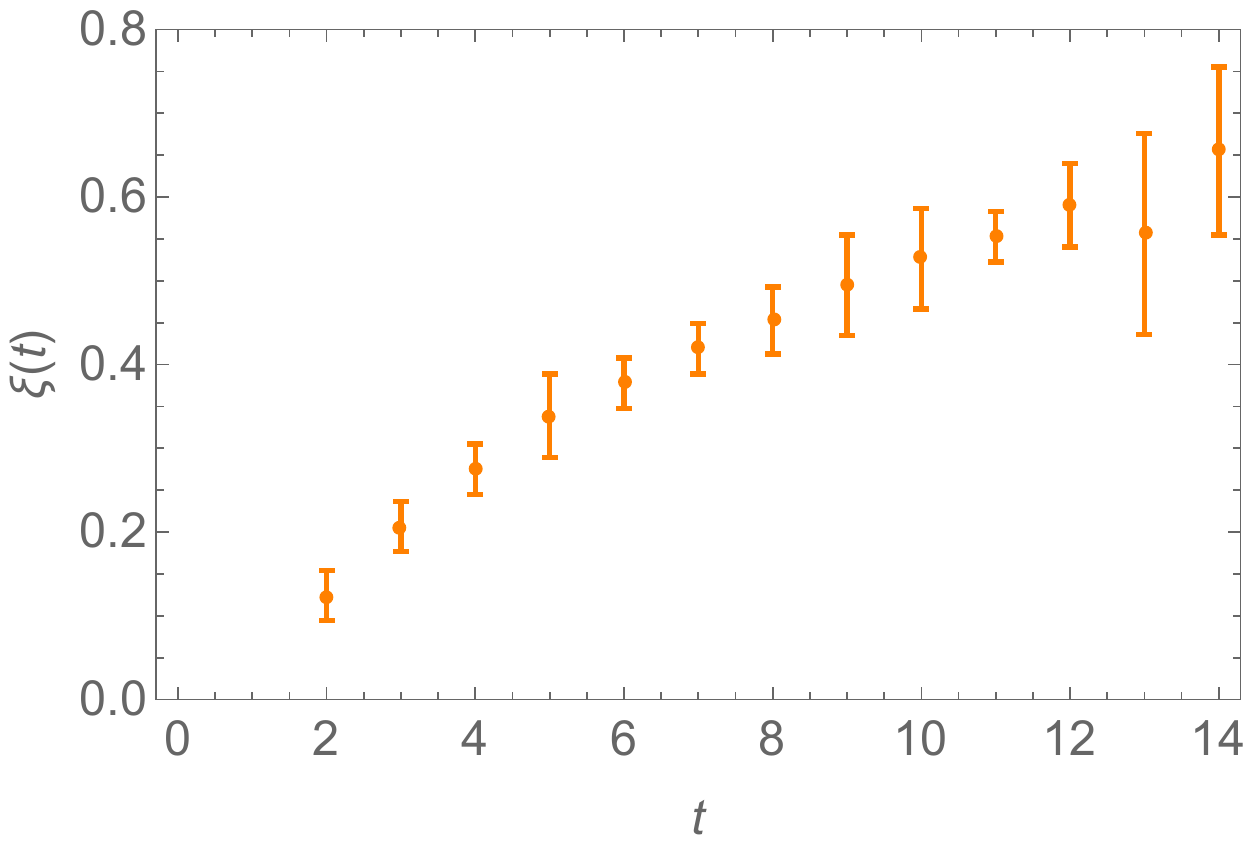}
\caption{Top: power spectrum of the Fourier transform of min-entropy across spatial cuts, normalized to one for all circuits. Note the narrowing of the Fourier transform with increasing $t$. Data are for a single system of length $5000$. Bottom: estimate of the correlation length $\xi$ extracted from the width of the Fourier peak.}
\label{ent_fourier}
\end{center}
\end{figure}

We now turn to fluctuations of the entanglement across spatial cuts. The prediction of Ref.~\cite{nrvh}, based on a mapping to the KPZ equation, is that the entanglement fluctuations are spatially correlated, with a correlation length $\xi(t) \sim t^{2/3}$, and that the width of the entanglement distribution scales as $t^{1/3}$ (i.e., the entanglement ``roughens''). The method used here works with an infinite system at a fixed depth, and enables one to address these spatial correlations. We find that the spatial correlations of entanglement do get longer-ranged in time, as their power spectrum clearly narrows in $k$-space (Fig.~\ref{ent_fourier}). The Fourier transform has a characteristic width, from which we can extract a correlation length that clearly grows sub-linearly with $t$. However, the correlation length remains short out to the latest times we can access, so we do not have the dynamic range to extract meaningful exponents.

The KPZ picture also predicts that the entanglement ``roughens'' with time, i.e., its standard deviation grows. This is consistent with what we see, although, again, the roughening is too weak to extract meaningful exponents.




%
%
%
%

\subsection{Trajectory Approach For Purity}

\begin{figure}[!b]
\begin{center}
\includegraphics[width = 0.45\textwidth]{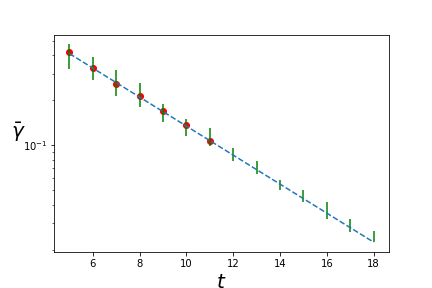}
\caption{Average purity of a random unitary circuit with $q=2$ computed by applying the quantum channel (red circles) and by the trajectory method (green bars) for $L=1000$ steps. Good agreement is found with the result $\bar\gamma = (4/5)^{t-1}$ from Ref.~\cite{Nahum2017} (dashed blue line).}
\label{fig:traj}
\end{center}
\end{figure}

Finally, we demonstrate the trajectory approach described in Section \ref{sec:traj}. To evaluate the purity using Eq.~\ref{eq:pure_traj} we evolve a pair of trajectories $s_{1:L}$ and $t_{1:L}$ with the transition probabilities given in Eq.~\eqref{eq:trans}. For a random unitary circuit with a purity that fluctuates with spatial position, evaluating the unaveraged purity at a point would involve averaging over many trajectories with the same set of gates. As a proof of principle we instead focus on the ensemble averaged purity, and average the fidelity $|\braket{\psi_{t_{1:l}}|\psi_{s_{1:l}}}|^2$ for $l=1,\ldots L$ with a trajectory of $L=1000$. In this case we can compare with the known exact result $\bar\gamma = (4/5)^{t-1}$ for random unitary circuits~\cite{Nahum2017}. Since we are now evolving vectors in the ancilla space rather than density matrices we can simulate deeper circuits. Fig.~\ref{fig:traj} shows the average purity for depths up to 18, comparing the trajectory method with the exact result, as well as with the density matrix approach for depths up to 12. Good agreement is found in all cases.

\subsection{Circuits with more structure}

The transfer-matrix method discussed here extends directly from random unitary circuits to any other type of circuit that can be decomposed into a ``brickwork'' arrangement of two-site gates. We discuss two examples here: circuits with a conservation law and translation-invariant circuits.

\subsubsection{Number-conserving circuits}

\begin{figure}[!b!t]
\begin{center}
\includegraphics[width = 0.45\textwidth]{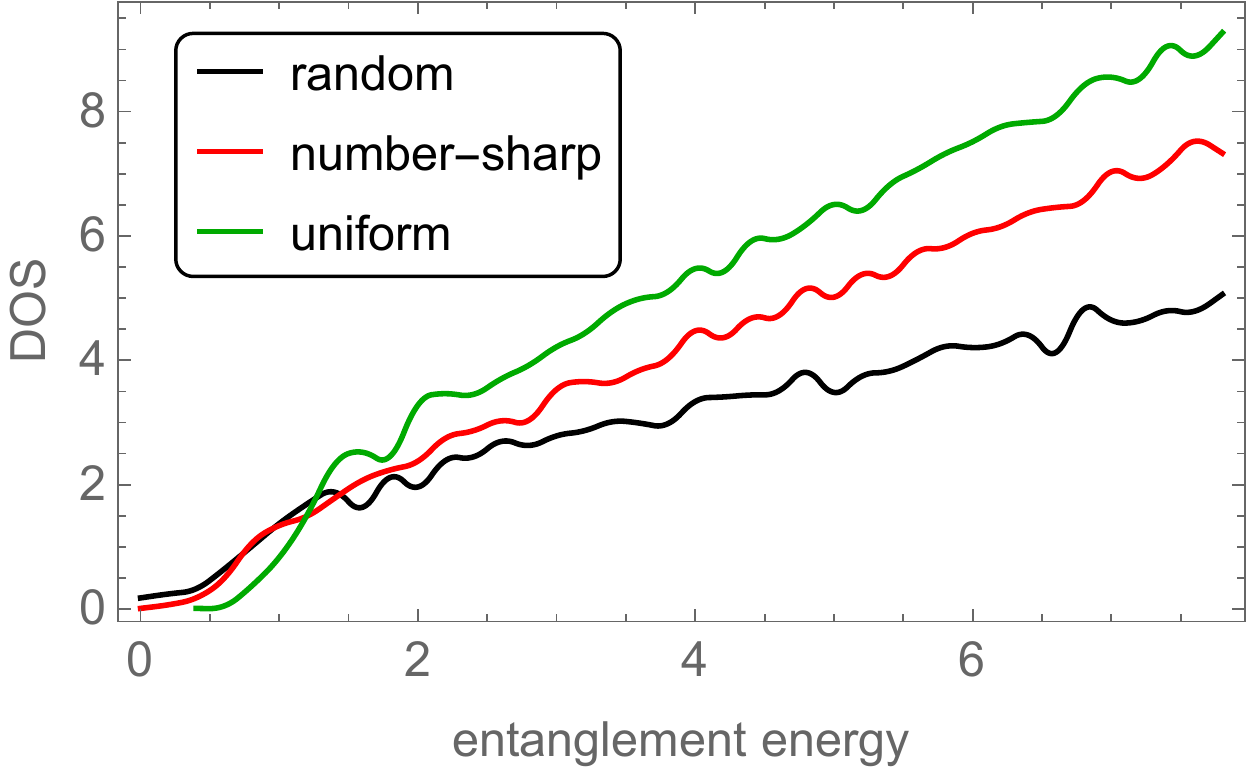}
\includegraphics[width = 0.45\textwidth]{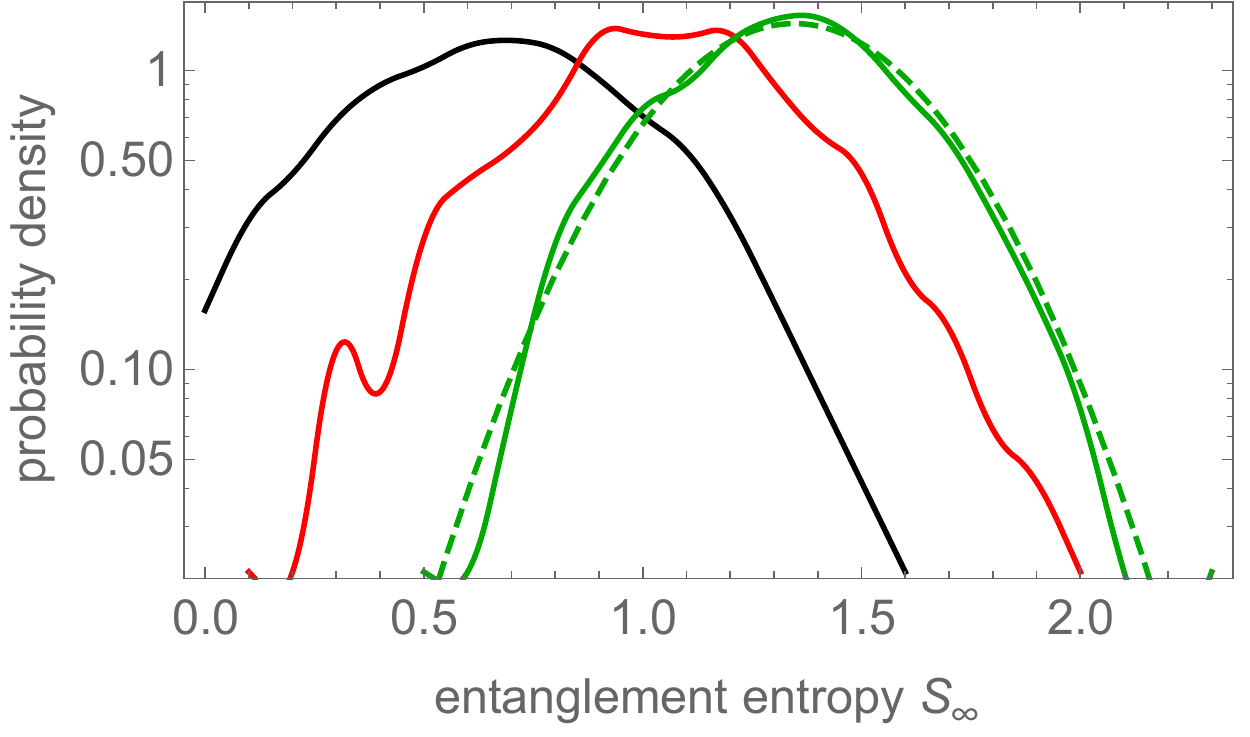}
\caption{\emph{Entanglement spectra for circuits with a conservation law}. Upper panel: behavior of the entanglement spectrum near its edge for three classes of initial states: random product states, random computational-basis bit-strings (``number-sharp''), and the nonrandom N\'eel state. Lower panel: fluctuations of the min-entropy for these three classes of states. All data are for depth $t = 12$, averaged over $3000$ samples. Only the initial N\'eel state approaches a Gaussian distribution.}
\label{conslaw1}
\end{center}
\end{figure}

As an illustrative example we now turn to circuits with a single conservation law, which we choose to be the number of $\uparrow$ spins in the computational basis~\cite{Khemani2017, rpv}. For a random circuit, all classes of product states are equivalent; however, circuits with a conservation law yield very different entanglement DOS depending on the initial state. Fig.~\ref{conslaw1} shows the results for three classes of initial states: random product states, random bit-strings in the computational basis, and a uniform N\'eel state. While the N\'eel state behaves analogously to the random unitary circuit, at least at these depths, we see that the other two types of product states give rise to very different entanglement spectra, with substantially higher DOS at low energies. The difference can be attributed to rare states with anomalously large weight on configurations with long strings of aligned spins, which do not entangle under number-conserving dynamics~\cite{rpv2019, yichen2019}.

\subsubsection{Translation-invariant circuits}

Next, we consider translation-invariant circuits, in which all the gates at a given time-step are identical. Gates could be the same at different time-steps (giving a Floquet system) or random at every time-step (giving a system with perfectly spatially correlated noise). We focus here on the former case. For concreteness we focus on the integrable Trotterization of the XXZ model that was recently introduced~\cite{Prosen_trotterization1, Prosen_trotterization2, Prosen_trotterization3}. This model is parameterized by two parameters $(\eta, \lambda)$, and the dynamics consists of repeated application of the two-site gate

\begin{equation}
U(\eta, \lambda) \equiv \left( \begin{array}{cccc} 1 & 0 & 0 & 0 \\ 0 & \frac{\sin \eta}{\sin (\eta + \lambda)} & \frac{\sin \lambda}{\sin(\eta + \lambda)} & 0 \\ 0 &  \frac{\sin \lambda}{\sin(\eta + \lambda)} &  \frac{\sin \eta}{\sin (\eta + \lambda)} & 0 \\ 0 & 0 & 0 & 1 \end{array} \right).
\end{equation}
When $\eta$ is imaginary and $\lambda$ is real, this is a Trotterized version of the Ising phase of the XXZ chain, with larger $\eta$ corresponding to larger easy-axis anisotropy.

\begin{figure}[tbh]
\begin{center}
\includegraphics[width = 0.45\textwidth]{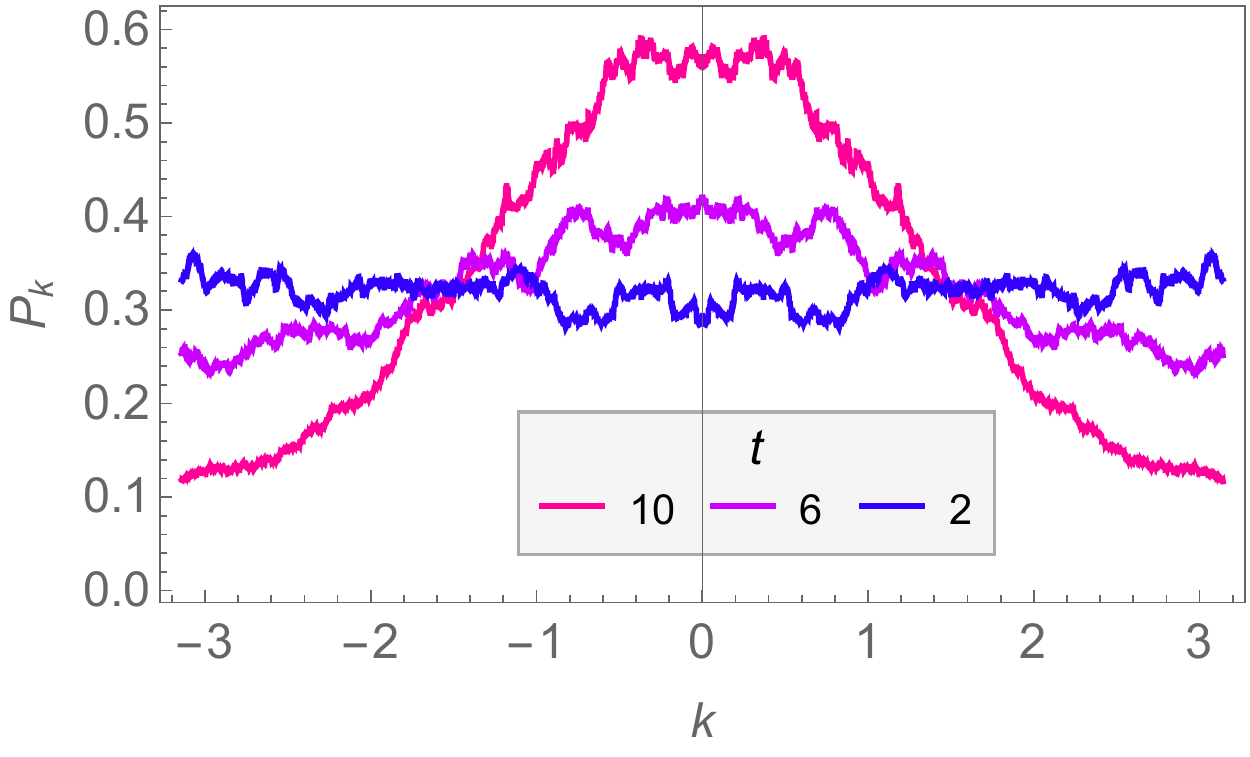}
\caption{Power spectrum of min-entropy fluctuations in the XXZ chain; the initial state is a random product state in the computational basis. As in random unitary circuits, the correlation length of the entanglement fluctuations grows in time.}
\label{xxzflucts}
\end{center}
\end{figure}

\begin{figure}[tbh]
\begin{center}
\includegraphics[width = 0.45\textwidth]{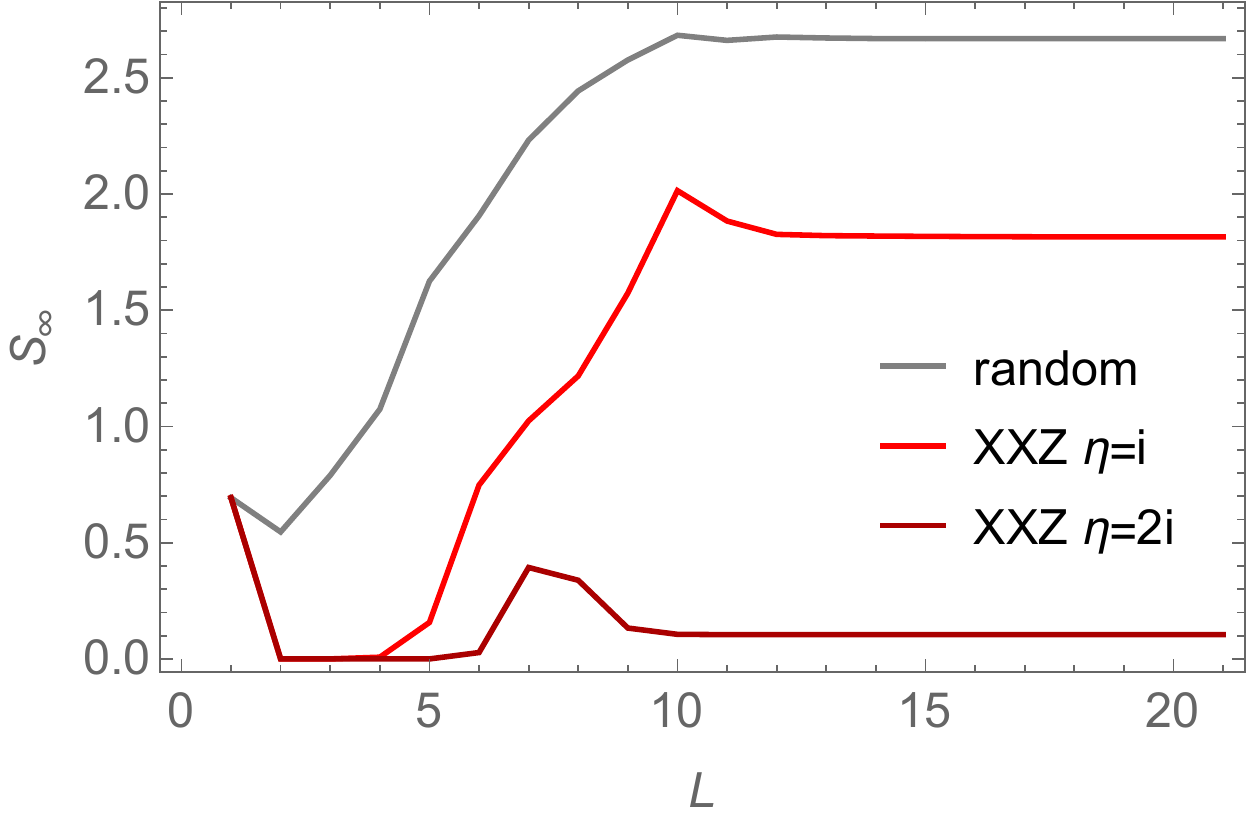}
\caption{Evolution of the min-entropy under the quantum channel, after $L$ steps, for depth $t = 12$. The data shown are for the Trotterized XXZ chain at two values of the anisotropy, and (for comparison) for a translation-invariant circuit with a randomly chosen gate. The integrable XXZ chain has non-monotonic entanglement growth, which is generically absent. In all cases the initial state is the N\'eel state $\uparrow \downarrow \uparrow \downarrow \ldots$.}
\label{convergence}
\end{center}
\end{figure}

\begin{figure}[tbh]
\begin{center}
\includegraphics[width = 0.45\textwidth]{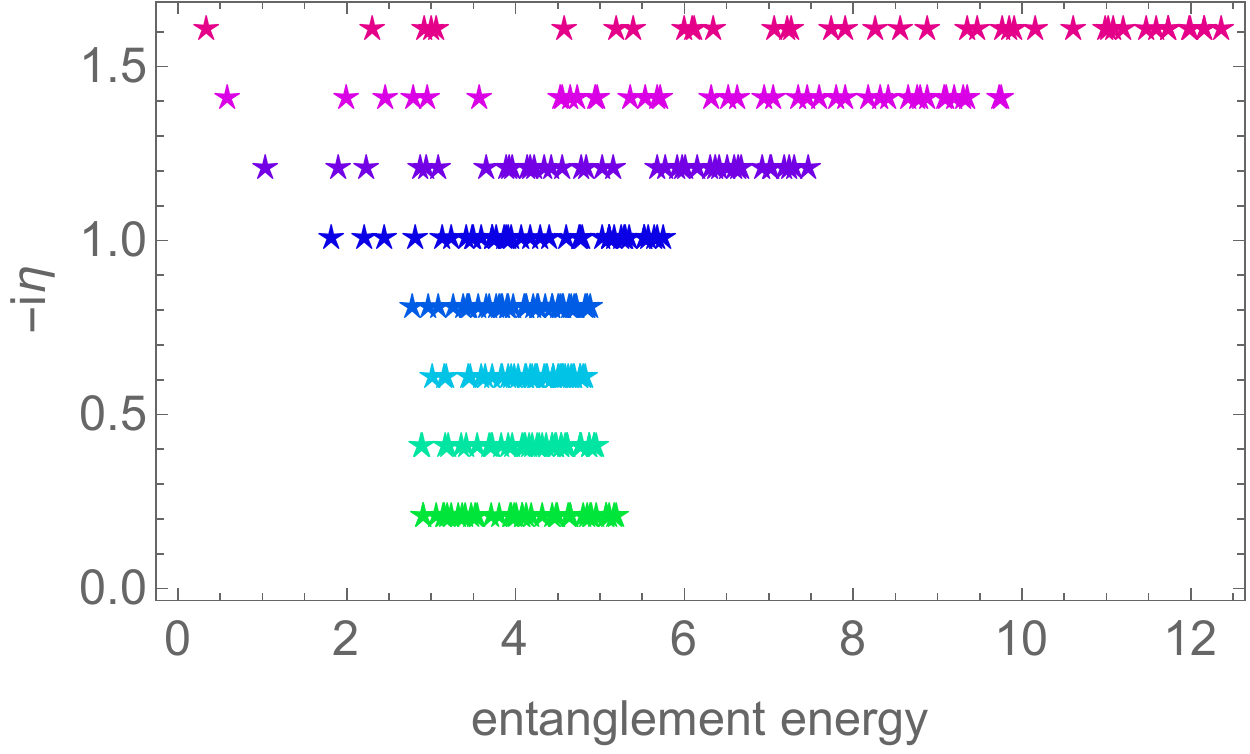}
\includegraphics[width = 0.45\textwidth]{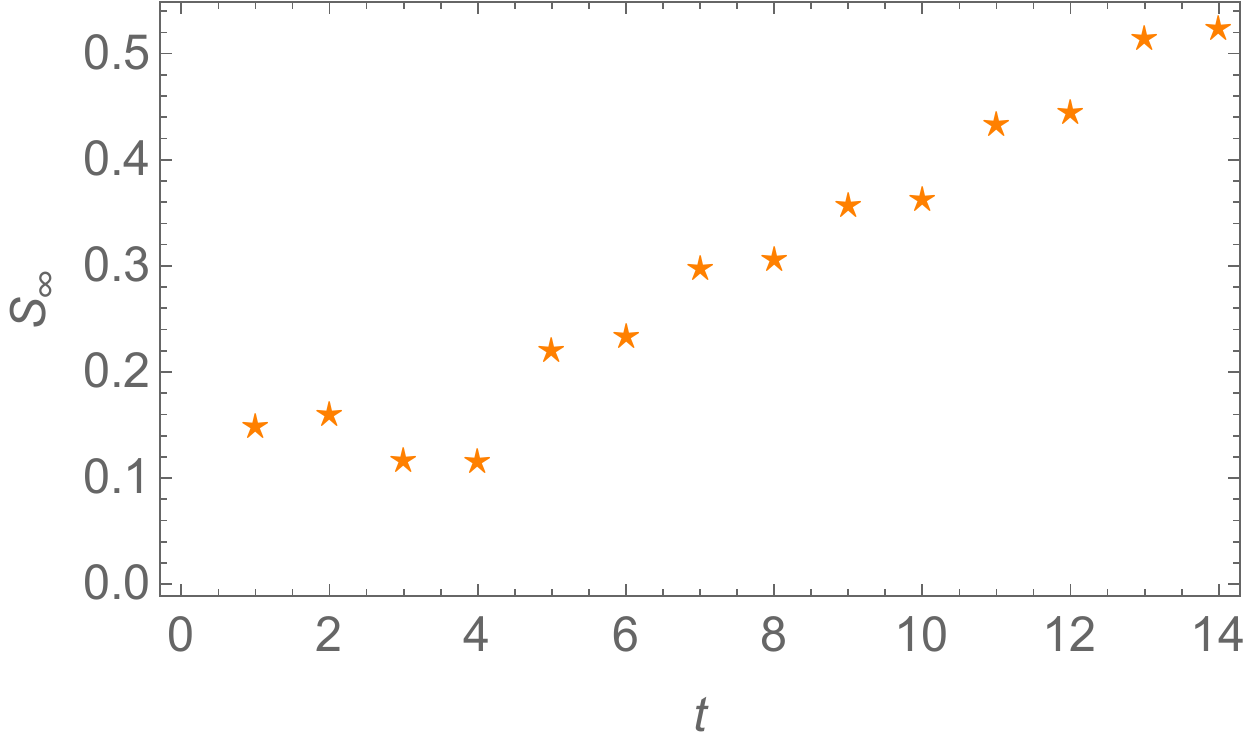}
\caption{Upper panel: entanglement spectra starting from the N\'eel state, under Trotterized XXZ dynamics, vs. anisotropy parameter $\eta$, at depth $12$. For large $\eta$ the dynamics freezes, so $S_\infty$ remains close to zero out to late times. Lower panel: time evolution of $S_\infty$ for a fixed anisotropy $\eta = 1.5 i$.}
\label{xxzent}
\end{center}
\end{figure}

One could consider the dynamics of entanglement for either random or homogeneous initial states. For random states, we expect (and find) spatial fluctuations of entanglement, which have a growing correlation length as in random unitary circuits (Fig.~\ref{xxzflucts}). At the accessible times, we are unable to extract any clear qualitative difference between the behavior of the correlation length in these integrable circuits and the random unitary case. For translation-invariant states (specifically the N\'eel state), the quantum channel converges to a definite steady state after a time interval on the order of the circuit depth (Fig.~\ref{convergence}). Comparing the transient behavior between the XXZ circuit and a (presumably nonintegrable) circuit consisting of tiling a random two-site gate, we see that the transient behavior of the integrable case is different: the min-entropy overshoots its steady-state value in the integrable case, but not in the random case. Fig.~\ref{xxzent} shows the lowest 40 entanglement energies at $t = 12$ as a function of the anisotropy parameter $\eta$; as one would expect, increasing the anisotropy slows down the growth of entanglement. This manifests itself as the top eigenvalue in the Schmidt spectrum drifting toward zero, leading to a large gap in the Schmidt spectrum. However, the min-entropy, starting from the N\'eel state, grows linearly out to the circuit depths we can access, with no signs of curvature. This is consistent with the intuitive picture of ballistic entanglement growth in integrable systems~\cite{alba_calabrese}, since quasiparticles move ballistically although spin dynamics is diffusive~\cite{lzp, dbd2, gv_superdiffusion, Prosen_trotterization3}.

An interesting quantitative difference between integrable dynamics and generic chaotic dynamics is that (for intermediate values of $\eta$) the entanglement spectrum stays ``narrow'' in the integrable case: Schmidt coefficients do not rapidly spread out over many decades the way they do under random unitary dynamics. This is consistent with the quasiparticle picture, which predicts that entanglement should spread with a characteristic quasiparticle velocity that does not depend on the R\'enyi index. How this picture extends to the case of strong anisotropy, in which the quasiparticles have a broad distribution of velocities, is an interesting open question.

\section{Conclusions}\label{conclusions}

In concluding, let us summarise the technical achievements of the quantum channel approach:

\begin{enumerate}
  \item Unitary circuits are presented directly as matrix product states in canonical form.
  \item The spectrum of the reduced density matrix is related to that of the ancilla states exposed by the diagonal cut.
  \item The resulting quantum channel allows us to work in the infinite width limit at finite depth.
  \item Analytical results are obtained for the kicked Ising model at the self-dual point using a simple graphical calculus, simpler than the approach of Ref.~\cite{Bertini:2018fbz}.
  \item The numerical evaluation of the channel may be improved by making a low rank approximation for the ancilla density matrix or by unraveling a quantum trajectory over the physical states.
\end{enumerate}

In this work, we benchmarked the quantum channel approach against exact results for the purity in random unitary circuits, and used it to compute the shape and fluctuations of the entanglement spectrum at low ``entanglement energies.'' With relatively little computational effort we were able to get converged results for depths $t = 14$ for the entanglement spectrum and $t = 18$ for the purity. We expect that there is room to optimize the algorithm and perform more resource-intensive computations, allowing us to access somewhat later times than we have in the present work. Our results support and extend earlier work~\cite{ccgp} showing that the entanglement spectrum has a bandwidth that grows rapidly in time, with a sharp onset. Under random unitary dynamics (with or without a conservation law), most of the Schmidt coefficients that are generated are exponentially small in circuit depth, and can therefore be truncated, allowing for efficient computation of the entanglement spectrum. We were able to compute the spatial fluctuations of entanglement for large systems, and finally for translation-invariant systems we obtained converged results for the entanglement spectrum by evolving the associated quantum channel to convergence.

A key distinction between the quantum channel approach and standard methods (such as time-evolving block decimation~\cite{vidal2003}) for evolving a matrix-product state is that the quantum channel approach constructs the entanglement spectrum without explicitly representing the physical wavefunction at time $t$. Our truncation and sampling schemes are also conceptually somewhat different from that in time-evolving block decimation~\cite{vidal2003, vidal2007}. Thus the quantum channel offers benefits if one wants to compute the entanglement spectrum (since it requires storing only the object of interest); however, it is not clear how one would apply our methods to compute observables in the physical (rather than ancilla) space.

A number of avenues for future work present themselves. The thermodynamic shape of the entanglement spectrum of a semi-infinite circuit is still not well understood beyond its coarsest features. At a technical level, applying our approach to the entanglement of a finite region will involve considering the quantum channel at different `times', instead of the stationary (distribution of the) ancilla density matrix. Finally, our analysis of the kicked Ising model shows that the soluble self-dual point arises simply from the relation Eq.~\eqref{eq:utilde}, and suggests a criterion for searching for more models in the same class.

\section{Acknowledgements}
This research was supported in part by the National Science Foundation under Grant No. NSF PHY-1748958. Support of EPSRC Grant No. EP/P034616/1 (AL) and NSF Grant No. DMR-1653271 (SG) is gratefully acknowledged. S.G. thanks P.-Y. Chang, X. Chen, V. Khemani, S. Parameswaran, J. Pixley, F. Pollmann, and T. Rakovszky for helpful discussions.


\begin{thebibliography}{41}
\expandafter\ifx\csname natexlab\endcsname\relax\def\natexlab#1{#1}\fi
\expandafter\ifx\csname bibnamefont\endcsname\relax
  \def\bibnamefont#1{#1}\fi
\expandafter\ifx\csname bibfnamefont\endcsname\relax
  \def\bibfnamefont#1{#1}\fi
\expandafter\ifx\csname citenamefont\endcsname\relax
  \def\citenamefont#1{#1}\fi
\expandafter\ifx\csname url\endcsname\relax
  \def\url#1{\texttt{#1}}\fi
\expandafter\ifx\csname urlprefix\endcsname\relax\def\urlprefix{URL }\fi
\providecommand{\bibinfo}[2]{#2}
\providecommand{\eprint}[2][]{\url{#2}}

\bibitem[{\citenamefont{Polkovnikov et~al.}(2011)\citenamefont{Polkovnikov,
  Sengupta, Silva, and Vengalattore}}]{polkovnikov_review}
\bibinfo{author}{\bibfnamefont{A.}~\bibnamefont{Polkovnikov}},
  \bibinfo{author}{\bibfnamefont{K.}~\bibnamefont{Sengupta}},
  \bibinfo{author}{\bibfnamefont{A.}~\bibnamefont{Silva}}, \bibnamefont{and}
  \bibinfo{author}{\bibfnamefont{M.}~\bibnamefont{Vengalattore}},
  \bibinfo{journal}{Rev. Mod. Phys.} \textbf{\bibinfo{volume}{83}},
  \bibinfo{pages}{863} (\bibinfo{year}{2011}),
  \urlprefix\url{https://link.aps.org/doi/10.1103/RevModPhys.83.863}.

\bibitem[{\citenamefont{Deutsch}(1991)}]{deutsch_eth}
\bibinfo{author}{\bibfnamefont{J.~M.} \bibnamefont{Deutsch}},
  \bibinfo{journal}{Phys. Rev. A} \textbf{\bibinfo{volume}{43}},
  \bibinfo{pages}{2046} (\bibinfo{year}{1991}),
  \urlprefix\url{https://link.aps.org/doi/10.1103/PhysRevA.43.2046}.

\bibitem[{\citenamefont{Srednicki}(1994)}]{srednicki_eth}
\bibinfo{author}{\bibfnamefont{M.}~\bibnamefont{Srednicki}},
  \bibinfo{journal}{Phys. Rev. E} \textbf{\bibinfo{volume}{50}},
  \bibinfo{pages}{888} (\bibinfo{year}{1994}),
  \urlprefix\url{https://link.aps.org/doi/10.1103/PhysRevE.50.888}.

\bibitem[{\citenamefont{Rigol et~al.}(2008)\citenamefont{Rigol, Dunjko, and
  Olshanii}}]{rigol2008thermalization}
\bibinfo{author}{\bibfnamefont{M.}~\bibnamefont{Rigol}},
  \bibinfo{author}{\bibfnamefont{V.}~\bibnamefont{Dunjko}}, \bibnamefont{and}
  \bibinfo{author}{\bibfnamefont{M.}~\bibnamefont{Olshanii}},
  \bibinfo{journal}{Nature} \textbf{\bibinfo{volume}{452}},
  \bibinfo{pages}{854} (\bibinfo{year}{2008}).

\bibitem[{\citenamefont{Deutsch}(2018)}]{deutsch_review}
\bibinfo{author}{\bibfnamefont{J.}~\bibnamefont{Deutsch}},
  \bibinfo{journal}{Reports on Progress in Physics}  (\bibinfo{year}{2018}).

\bibitem[{\citenamefont{Calabrese and Cardy}(2005)}]{calabrese2005evolution}
\bibinfo{author}{\bibfnamefont{P.}~\bibnamefont{Calabrese}} \bibnamefont{and}
  \bibinfo{author}{\bibfnamefont{J.}~\bibnamefont{Cardy}},
  \bibinfo{journal}{Journal of Statistical Mechanics: Theory and Experiment}
  \textbf{\bibinfo{volume}{2005}}, \bibinfo{pages}{P04010}
  (\bibinfo{year}{2005}).

\bibitem[{\citenamefont{Kim and Huse}(2013)}]{Kim:2013aa}
\bibinfo{author}{\bibfnamefont{H.}~\bibnamefont{Kim}} \bibnamefont{and}
  \bibinfo{author}{\bibfnamefont{D.~A.} \bibnamefont{Huse}},
  \bibinfo{journal}{Phys. Rev. Lett.} \textbf{\bibinfo{volume}{111}},
  \bibinfo{pages}{127205} (\bibinfo{year}{2013}),
  \urlprefix\url{https://link.aps.org/doi/10.1103/PhysRevLett.111.127205}.

\bibitem[{\citenamefont{Nahum et~al.}(2017)\citenamefont{Nahum, Ruhman, Vijay,
  and Haah}}]{nrvh}
\bibinfo{author}{\bibfnamefont{A.}~\bibnamefont{Nahum}},
  \bibinfo{author}{\bibfnamefont{J.}~\bibnamefont{Ruhman}},
  \bibinfo{author}{\bibfnamefont{S.}~\bibnamefont{Vijay}}, \bibnamefont{and}
  \bibinfo{author}{\bibfnamefont{J.}~\bibnamefont{Haah}},
  \bibinfo{journal}{Phys. Rev. X} \textbf{\bibinfo{volume}{7}},
  \bibinfo{pages}{031016} (\bibinfo{year}{2017}),
  \urlprefix\url{https://link.aps.org/doi/10.1103/PhysRevX.7.031016}.

\bibitem[{\citenamefont{von Keyserlingk et~al.}(2018)\citenamefont{von
  Keyserlingk, Rakovszky, Pollmann, and Sondhi}}]{Keyserlingk2017}
\bibinfo{author}{\bibfnamefont{C.~W.} \bibnamefont{von Keyserlingk}},
  \bibinfo{author}{\bibfnamefont{T.}~\bibnamefont{Rakovszky}},
  \bibinfo{author}{\bibfnamefont{F.}~\bibnamefont{Pollmann}}, \bibnamefont{and}
  \bibinfo{author}{\bibfnamefont{S.~L.} \bibnamefont{Sondhi}},
  \bibinfo{journal}{Phys. Rev. X} \textbf{\bibinfo{volume}{8}},
  \bibinfo{pages}{021013} (\bibinfo{year}{2018}),
  \urlprefix\url{https://link.aps.org/doi/10.1103/PhysRevX.8.021013}.

\bibitem[{\citenamefont{Nahum et~al.}(2018)\citenamefont{Nahum, Vijay, and
  Haah}}]{Nahum2017}
\bibinfo{author}{\bibfnamefont{A.}~\bibnamefont{Nahum}},
  \bibinfo{author}{\bibfnamefont{S.}~\bibnamefont{Vijay}}, \bibnamefont{and}
  \bibinfo{author}{\bibfnamefont{J.}~\bibnamefont{Haah}},
  \bibinfo{journal}{Phys. Rev. X} \textbf{\bibinfo{volume}{8}},
  \bibinfo{pages}{021014} (\bibinfo{year}{2018}),
  \urlprefix\url{https://link.aps.org/doi/10.1103/PhysRevX.8.021014}.

\bibitem[{\citenamefont{{\v{Z}}nidari{\v{c}}
  et~al.}(2008)\citenamefont{{\v{Z}}nidari{\v{c}}, Prosen, and
  Prelov{\v{s}}ek}}]{Znidaric:2008aa}
\bibinfo{author}{\bibfnamefont{M.}~\bibnamefont{{\v{Z}}nidari{\v{c}}}},
  \bibinfo{author}{\bibfnamefont{T.}~\bibnamefont{Prosen}}, \bibnamefont{and}
  \bibinfo{author}{\bibfnamefont{P.}~\bibnamefont{Prelov{\v{s}}ek}},
  \bibinfo{journal}{Physical Review B} \textbf{\bibinfo{volume}{77}},
  \bibinfo{pages}{064426} (\bibinfo{year}{2008}).

\bibitem[{\citenamefont{Rakovszky et~al.}(2019)\citenamefont{Rakovszky,
  Pollmann, and von Keyserlingk}}]{rpv2019}
\bibinfo{author}{\bibfnamefont{T.}~\bibnamefont{Rakovszky}},
  \bibinfo{author}{\bibfnamefont{F.}~\bibnamefont{Pollmann}}, \bibnamefont{and}
  \bibinfo{author}{\bibfnamefont{C.}~\bibnamefont{von Keyserlingk}},
  \bibinfo{journal}{arXiv preprint arXiv:1901.10502}  (\bibinfo{year}{2019}).

\bibitem[{\citenamefont{Huang}(2019)}]{yichen2019}
\bibinfo{author}{\bibfnamefont{Y.}~\bibnamefont{Huang}},
  \bibinfo{journal}{arXiv preprint arXiv:1902.00977}  (\bibinfo{year}{2019}).

\bibitem[{\citenamefont{Li and Haldane}(2008)}]{ent_spectrum}
\bibinfo{author}{\bibfnamefont{H.}~\bibnamefont{Li}} \bibnamefont{and}
  \bibinfo{author}{\bibfnamefont{F.~D.~M.} \bibnamefont{Haldane}},
  \bibinfo{journal}{Phys. Rev. Lett.} \textbf{\bibinfo{volume}{101}},
  \bibinfo{pages}{010504} (\bibinfo{year}{2008}),
  \urlprefix\url{https://link.aps.org/doi/10.1103/PhysRevLett.101.010504}.

\bibitem[{\citenamefont{Pollmann and Moore}(2010)}]{pollmann2010}
\bibinfo{author}{\bibfnamefont{F.}~\bibnamefont{Pollmann}} \bibnamefont{and}
  \bibinfo{author}{\bibfnamefont{J.~E.} \bibnamefont{Moore}},
  \bibinfo{journal}{New Journal of Physics} \textbf{\bibinfo{volume}{12}},
  \bibinfo{pages}{025006} (\bibinfo{year}{2010}).

\bibitem[{\citenamefont{Ho and Abanin}(2017)}]{ho2017}
\bibinfo{author}{\bibfnamefont{W.~W.} \bibnamefont{Ho}} \bibnamefont{and}
  \bibinfo{author}{\bibfnamefont{D.~A.} \bibnamefont{Abanin}},
  \bibinfo{journal}{Phys. Rev. B} \textbf{\bibinfo{volume}{95}},
  \bibinfo{pages}{094302} (\bibinfo{year}{2017}),
  \urlprefix\url{https://link.aps.org/doi/10.1103/PhysRevB.95.094302}.

\bibitem[{\citenamefont{Mezei and Stanford}(2017)}]{mezei2017entanglement}
\bibinfo{author}{\bibfnamefont{M.}~\bibnamefont{Mezei}} \bibnamefont{and}
  \bibinfo{author}{\bibfnamefont{D.}~\bibnamefont{Stanford}},
  \bibinfo{journal}{Journal of High Energy Physics}
  \textbf{\bibinfo{volume}{2017}}, \bibinfo{pages}{65} (\bibinfo{year}{2017}).

\bibitem[{\citenamefont{Zhou and Nahum}(2018)}]{tianci}
\bibinfo{author}{\bibfnamefont{T.}~\bibnamefont{Zhou}} \bibnamefont{and}
  \bibinfo{author}{\bibfnamefont{A.}~\bibnamefont{Nahum}},
  \bibinfo{journal}{arXiv preprint arXiv:1804.09737}  (\bibinfo{year}{2018}).

\bibitem[{\citenamefont{Bertini
  et~al.}(2018{\natexlab{a}})\citenamefont{Bertini, Kos, and
  Prosen}}]{Bertini:2018aa}
\bibinfo{author}{\bibfnamefont{B.}~\bibnamefont{Bertini}},
  \bibinfo{author}{\bibfnamefont{P.}~\bibnamefont{Kos}}, \bibnamefont{and}
  \bibinfo{author}{\bibfnamefont{T.}~\bibnamefont{Prosen}},
  \bibinfo{journal}{arXiv preprint arXiv:1805.00931}
  (\bibinfo{year}{2018}{\natexlab{a}}).

\bibitem[{\citenamefont{Bertini
  et~al.}(2018{\natexlab{b}})\citenamefont{Bertini, Kos, and
  Prosen}}]{Bertini:2018fbz}
\bibinfo{author}{\bibfnamefont{B.}~\bibnamefont{Bertini}},
  \bibinfo{author}{\bibfnamefont{P.}~\bibnamefont{Kos}}, \bibnamefont{and}
  \bibinfo{author}{\bibfnamefont{T.}~\bibnamefont{Prosen}}
  (\bibinfo{year}{2018}{\natexlab{b}}), \eprint{1812.05090}.

\bibitem[{\citenamefont{Chang et~al.}(2018)\citenamefont{Chang, Chen,
  Gopalakrishnan, and Pixley}}]{ccgp}
\bibinfo{author}{\bibfnamefont{P.-Y.} \bibnamefont{Chang}},
  \bibinfo{author}{\bibfnamefont{X.}~\bibnamefont{Chen}},
  \bibinfo{author}{\bibfnamefont{S.}~\bibnamefont{Gopalakrishnan}},
  \bibnamefont{and} \bibinfo{author}{\bibfnamefont{J.}~\bibnamefont{Pixley}},
  \bibinfo{journal}{arXiv preprint arXiv:1811.00029}  (\bibinfo{year}{2018}).

\bibitem[{\citenamefont{Carmichael}(2009)}]{carmichael_book}
\bibinfo{author}{\bibfnamefont{H.}~\bibnamefont{Carmichael}},
  \emph{\bibinfo{title}{An open systems approach to quantum optics: lectures
  presented at the Universit{\'e} Libre de Bruxelles, October 28 to November 4,
  1991}}, vol.~\bibinfo{volume}{18} (\bibinfo{publisher}{Springer Science \&
  Business Media}, \bibinfo{year}{2009}).

\bibitem[{\citenamefont{Vidal}(2007)}]{vidal2007}
\bibinfo{author}{\bibfnamefont{G.}~\bibnamefont{Vidal}},
  \bibinfo{journal}{Phys. Rev. Lett.} \textbf{\bibinfo{volume}{98}},
  \bibinfo{pages}{070201} (\bibinfo{year}{2007}),
  \urlprefix\url{https://link.aps.org/doi/10.1103/PhysRevLett.98.070201}.

\bibitem[{\citenamefont{Perez-Garcia et~al.}(2006)\citenamefont{Perez-Garcia,
  Verstraete, Wolf, and Cirac}}]{Perez-Garcia:2006aa}
\bibinfo{author}{\bibfnamefont{D.}~\bibnamefont{Perez-Garcia}},
  \bibinfo{author}{\bibfnamefont{F.}~\bibnamefont{Verstraete}},
  \bibinfo{author}{\bibfnamefont{M.~M.} \bibnamefont{Wolf}}, \bibnamefont{and}
  \bibinfo{author}{\bibfnamefont{J.~I.} \bibnamefont{Cirac}},
  \bibinfo{journal}{arXiv preprint quant-ph/0608197}  (\bibinfo{year}{2006}).

\bibitem[{\citenamefont{Schollw{\"o}ck}(2011)}]{Schollwock:2011aa}
\bibinfo{author}{\bibfnamefont{U.}~\bibnamefont{Schollw{\"o}ck}},
  \bibinfo{journal}{Annals of Physics} \textbf{\bibinfo{volume}{326}},
  \bibinfo{pages}{96} (\bibinfo{year}{2011}).

\bibitem[{\citenamefont{Or{\'u}s}(2014)}]{Orus:2014aa}
\bibinfo{author}{\bibfnamefont{R.}~\bibnamefont{Or{\'u}s}},
  \bibinfo{journal}{Annals of Physics} \textbf{\bibinfo{volume}{349}},
  \bibinfo{pages}{117} (\bibinfo{year}{2014}).

\bibitem[{\citenamefont{Lindblad}(1976)}]{lindblad1976generators}
\bibinfo{author}{\bibfnamefont{G.}~\bibnamefont{Lindblad}},
  \bibinfo{journal}{Communications in Mathematical Physics}
  \textbf{\bibinfo{volume}{48}}, \bibinfo{pages}{119} (\bibinfo{year}{1976}).

\bibitem[{\citenamefont{Gopalakrishnan and Zakirov}(2018)}]{gz2018}
\bibinfo{author}{\bibfnamefont{S.}~\bibnamefont{Gopalakrishnan}}
  \bibnamefont{and} \bibinfo{author}{\bibfnamefont{B.}~\bibnamefont{Zakirov}},
  \bibinfo{journal}{arXiv preprint arXiv:1802.07729}  (\bibinfo{year}{2018}).

\bibitem[{\citenamefont{Gopalakrishnan}(2018)}]{sg_og_2018}
\bibinfo{author}{\bibfnamefont{S.}~\bibnamefont{Gopalakrishnan}},
  \bibinfo{journal}{Phys. Rev. B} \textbf{\bibinfo{volume}{98}},
  \bibinfo{pages}{060302} (\bibinfo{year}{2018}),
  \urlprefix\url{https://link.aps.org/doi/10.1103/PhysRevB.98.060302}.

\bibitem[{Note1()}]{Note1}
\bibinfo{note}{The exponent $t-1$ here is one less than in \cite
  {Nahum2017} because our partition of the system lies between the two gates in
  the top layer of the circuit, which therefore leaves the entanglement
  unaffected.}

\bibitem[{\citenamefont{Chen and Ludwig}(2017)}]{Chen:2017aa}
\bibinfo{author}{\bibfnamefont{X.}~\bibnamefont{Chen}} \bibnamefont{and}
  \bibinfo{author}{\bibfnamefont{A.~W.} \bibnamefont{Ludwig}},
  \bibinfo{journal}{arXiv preprint arXiv:1710.02686}  (\bibinfo{year}{2017}).

\bibitem[{\citenamefont{Khemani et~al.}(2018)\citenamefont{Khemani, Vishwanath,
  and Huse}}]{Khemani2017}
\bibinfo{author}{\bibfnamefont{V.}~\bibnamefont{Khemani}},
  \bibinfo{author}{\bibfnamefont{A.}~\bibnamefont{Vishwanath}},
  \bibnamefont{and} \bibinfo{author}{\bibfnamefont{D.~A.} \bibnamefont{Huse}},
  \bibinfo{journal}{Phys. Rev. X} \textbf{\bibinfo{volume}{8}},
  \bibinfo{pages}{031057} (\bibinfo{year}{2018}),
  \urlprefix\url{https://link.aps.org/doi/10.1103/PhysRevX.8.031057}.

\bibitem[{\citenamefont{Rakovszky et~al.}(2018)\citenamefont{Rakovszky,
  Pollmann, and von Keyserlingk}}]{rpv}
\bibinfo{author}{\bibfnamefont{T.}~\bibnamefont{Rakovszky}},
  \bibinfo{author}{\bibfnamefont{F.}~\bibnamefont{Pollmann}}, \bibnamefont{and}
  \bibinfo{author}{\bibfnamefont{C.~W.} \bibnamefont{von Keyserlingk}},
  \bibinfo{journal}{Phys. Rev. X} \textbf{\bibinfo{volume}{8}},
  \bibinfo{pages}{031058} (\bibinfo{year}{2018}),
  \urlprefix\url{https://link.aps.org/doi/10.1103/PhysRevX.8.031058}.

\bibitem[{\citenamefont{Vanicat et~al.}(2018)\citenamefont{Vanicat, Zadnik, and
  Prosen}}]{Prosen_trotterization1}
\bibinfo{author}{\bibfnamefont{M.}~\bibnamefont{Vanicat}},
  \bibinfo{author}{\bibfnamefont{L.}~\bibnamefont{Zadnik}}, \bibnamefont{and}
  \bibinfo{author}{\bibfnamefont{T.~c.~v.} \bibnamefont{Prosen}},
  \bibinfo{journal}{Phys. Rev. Lett.} \textbf{\bibinfo{volume}{121}},
  \bibinfo{pages}{030606} (\bibinfo{year}{2018}),
  \urlprefix\url{https://link.aps.org/doi/10.1103/PhysRevLett.121.030606}.

\bibitem[{\citenamefont{Ljubotina
  et~al.}(2019{\natexlab{a}})\citenamefont{Ljubotina, Zadnik, and
  Prosen}}]{Prosen_trotterization2}
\bibinfo{author}{\bibfnamefont{M.}~\bibnamefont{Ljubotina}},
  \bibinfo{author}{\bibfnamefont{L.}~\bibnamefont{Zadnik}}, \bibnamefont{and}
  \bibinfo{author}{\bibfnamefont{T.}~\bibnamefont{Prosen}},
  \bibinfo{journal}{arXiv preprint arXiv:1901.05398}
  (\bibinfo{year}{2019}{\natexlab{a}}).

\bibitem[{\citenamefont{Ljubotina
  et~al.}(2019{\natexlab{b}})\citenamefont{Ljubotina, Znidaric, and
  Prosen}}]{Prosen_trotterization3}
\bibinfo{author}{\bibfnamefont{M.}~\bibnamefont{Ljubotina}},
  \bibinfo{author}{\bibfnamefont{M.}~\bibnamefont{Znidaric}}, \bibnamefont{and}
  \bibinfo{author}{\bibfnamefont{T.}~\bibnamefont{Prosen}},
  \bibinfo{journal}{arXiv preprint arXiv:1903.01329}
  (\bibinfo{year}{2019}{\natexlab{b}}).

\bibitem[{\citenamefont{Alba and Calabrese}(2017)}]{alba_calabrese}
\bibinfo{author}{\bibfnamefont{V.}~\bibnamefont{Alba}} \bibnamefont{and}
  \bibinfo{author}{\bibfnamefont{P.}~\bibnamefont{Calabrese}},
  \bibinfo{journal}{Proceedings of the National Academy of Sciences}
  \textbf{\bibinfo{volume}{114}}, \bibinfo{pages}{7947} (\bibinfo{year}{2017}).

\bibitem[{\citenamefont{Ljubotina et~al.}(2017)\citenamefont{Ljubotina, {\v
  Z}nidari{\v c}, and Prosen}}]{lzp}
\bibinfo{author}{\bibfnamefont{M.}~\bibnamefont{Ljubotina}},
  \bibinfo{author}{\bibfnamefont{M.}~\bibnamefont{{\v Z}nidari{\v c}}},
  \bibnamefont{and} \bibinfo{author}{\bibfnamefont{T.}~\bibnamefont{Prosen}},
  \bibinfo{journal}{Nature Communications} \textbf{\bibinfo{volume}{8}},
  \bibinfo{pages}{16117 EP } (\bibinfo{year}{2017}),
  \urlprefix\url{http://dx.doi.org/10.1038/ncomms16117}.

\bibitem[{\citenamefont{De~Nardis et~al.}(2018)\citenamefont{De~Nardis,
  Bernard, and Doyon}}]{dbd2}
\bibinfo{author}{\bibfnamefont{J.}~\bibnamefont{De~Nardis}},
  \bibinfo{author}{\bibfnamefont{D.}~\bibnamefont{Bernard}}, \bibnamefont{and}
  \bibinfo{author}{\bibfnamefont{B.}~\bibnamefont{Doyon}},
  \bibinfo{journal}{arXiv preprint arXiv:1812.00767}  (\bibinfo{year}{2018}).

\bibitem[{\citenamefont{Gopalakrishnan and Vasseur}(2018)}]{gv_superdiffusion}
\bibinfo{author}{\bibfnamefont{S.}~\bibnamefont{Gopalakrishnan}}
  \bibnamefont{and} \bibinfo{author}{\bibfnamefont{R.}~\bibnamefont{Vasseur}},
  \bibinfo{journal}{arXiv preprint arXiv:1812.02701}  (\bibinfo{year}{2018}).

\bibitem[{\citenamefont{Vidal}(2003)}]{vidal2003}
\bibinfo{author}{\bibfnamefont{G.}~\bibnamefont{Vidal}},
  \bibinfo{journal}{Phys. Rev. Lett.} \textbf{\bibinfo{volume}{91}},
  \bibinfo{pages}{147902} (\bibinfo{year}{2003}),
  \urlprefix\url{https://link.aps.org/doi/10.1103/PhysRevLett.91.147902}.

\end{thebibliography}

\end{document}